\documentclass[sigconf]{acmart}

\usepackage{booktabs} % For formal tables
\usepackage{multirow} % to combine the cells vertically in tables
\usepackage{graphicx} % Required for inserting images
\usepackage{cleveref}
\usepackage{subcaption}
\usepackage[xcdraw]{colortbl}
\usepackage{footnote} % footnote packages for tables
\usepackage{footmisc} % footnote packages for tables
\usepackage{threeparttable} % for adding notes under the table
\usepackage{makecell}
\usepackage{pdfpages}
\usepackage[breakable, skins]{tcolorbox}
\usepackage{adjustbox}
\usepackage{natbib}
\usepackage{gensymb}
\usepackage[normalem]{ulem}
\usepackage{dblfloatfix}
\usepackage{dblfloatfix}
\usepackage[absolute,showboxes]{textpos}

\tcbset{arc=0mm,size=fbox,attach title to upper={\ ---\ },coltitle=black}

\renewcommand\footnotetextcopyrightpermission[1]{} % removes footnote with conference info
\setcopyright{none}
\acmConference[WWW '24]{Proceedings of the ACM Web Conference 2024}{May 13--17, 2024}{Singapore, Singapore}
\title{A Multifaceted Look at Starlink Performance}
\author{Nitinder Mohan}
\authornote{Authors contributed equally to this research.}
\affiliation{%
   \institution{Technical University of Munich}
   \country{Germany}}
\author{Andrew E. Ferguson}
\authornotemark[1]
\affiliation{%
   \institution{The University of Edinburgh}
   \country{United Kingdom}}
\author{Hendrik Cech}
\authornotemark[1]
\affiliation{%
   \institution{Technical University of Munich}
   \country{Germany}}
\author{Rohan Bose}
\affiliation{%
   \institution{Technical University of Munich}
   \country{Germany}}
\author{Prakita Rayyan Renatin}
\affiliation{%
   \institution{Technical University of Munich}
   \country{Germany}}
\author{Mahesh K. Marina}
\affiliation{%
   \institution{The University of Edinburgh}
   \country{United Kingdom}}
\author{Jörg Ott}
\affiliation{%
   \institution{Technical University of Munich}
   \country{Germany}}
\crefformat{section}{\S#2#1#3}
\crefformat{subsection}{\S#2#1#3}
\crefformat{subsubsection}{\S#2#1#3}
\crefrangeformat{section}{\S\S#3#1#4 to~#5#2#6}
\crefmultiformat{section}{\S\S#2#1#3}{ and~#2#1#3}{, #2#1#3}{ and~#2#1#3}

\newcommand{\headline}[1]{\smallskip\noindent\textbf{\textit{#1.}}}
\newcommand{\stddev}[1]{{\footnotesize$\pm$#1}}

\begin{document}

\begin{abstract}
    % \textcolor{red}{NM: To be handled}
    In recent years, Low-Earth Orbit (LEO) mega-constellations have ushered in a new era for ubiquitous Internet access. 
    % to access the Internet.
    The Starlink network from SpaceX stands out as the only commercial LEO network with over 2M+ customers and more than 4000 operational satellites.
    %
    % Prior works have analyzed Starlink's performance in small-scale, targeted studies.
    %
    In this paper, we conduct a first-of-its-kind extensive multi-faceted analysis of Starlink performance leveraging several measurement sources.
    % towards an encompassing view spanning its performance globally as well as its internal ``bent-pipe'' behavior.        
    %
    First, based on 19.2M crowdsourced M-Lab speed tests from 34 countries since 2021, we analyze Starlink global performance relative to terrestrial cellular networks.
    % and performance anomalies in insufficiently covered regions.
    %
    Second, we examine Starlink's ability to support real-time latency and bandwidth-critical applications by analyzing the performance of (i) Zoom conferencing, and (ii) Luna cloud gaming, comparing it to 5G and fiber. 
    Third, we perform measurements from Starlink-enabled RIPE Atlas probes to shed light on the last-mile access and other factors affecting its performance.
    % in regions with different satellite coverage. 
    % complement our crowdsourced data and to get an insight into the behavior of Starlink's bent-pipe in different regions.
    %
    Finally, we conduct controlled experiments from Starlink dishes in two countries and analyze the impact of globally synchronized ``15-second reconfiguration intervals'' of the satellite links that cause substantial latency and throughput variations.
    %
    % Finally, we highlight Starlink's capabilities to support real-time web-based latency and bandwidth-critical applications by analyzing the performance of (i) cloud gaming, and (ii) Zoom video conferencing, comparing it to 5G and terrestrial WiFi networks. 
    %
    Our unique analysis 
    % provides revealing insights on global Starlink functionality and 
    paints the most comprehensive picture of Starlink's global and last-mile performance to date. 
 %   We reflect on whether Starlink will keep pace with the emerging and ever-more demanding networked applications and growth in the number of users. 
\end{abstract}

    % assemble a global view of the network's current performance, and of its last-mile characteristics:

%% To upload
% In recent years, Low-Earth Orbit (LEO) mega-constellations have ushered in a new era for ubiquitous Internet access. 
% The Starlink network from SpaceX stands out as the only commercial LEO network with over 2M+ customers and more than 4000 operational satellites. In this paper, we conduct a first-of-its-kind extensive multi-faceted analysis of Starlink performance leveraging several measurement sources. First, based on 19.2M crowdsourced M-Lab speed tests from 34 countries since 2021, we analyze Starlink global performance relative to terrestrial cellular networks. Second, we examine Starlink's ability to support real-time latency and bandwidth-critical applications by analyzing the performance of (i) Zoom conferencing, and (ii) Luna cloud gaming, comparing it to 5G and fiber. Third, we perform measurements from Starlink-enabled RIPE Atlas probes to shed light on the last-mile access and other factors affecting its performance.Finally, we conduct controlled experiments from Starlink dishes in two countries and analyze the impact of globally synchronized ``15-second reconfiguration intervals'' of the satellite links that cause substantial latency and throughput variations. Our unique analysis paints the most comprehensive picture of Starlink's global and last-mile performance to date.

\setlength{\TPHorizModule}{\paperwidth}
\setlength{\TPVertModule}{\paperheight}
\TPMargin{5pt}
\begin{textblock}{0.8}(0.1,0.02)
    \noindent
   %  \footnotesize
    If you cite this paper, please use the The Web Conference reference:
    "Nitinder Mohan, Andrew Fergusson, Hendrik Cech, Rohan Bose, Prakita Rayyan Renatin, Mahesh K. Marina, Jörg Ott.
    2024. A Multifaceted Look at Starlink Performance. In Proceedings of the \textit{ACM Web Conference 2024 (WWW ’24)}, May 13–17, 2024, Singapore, Singapore. ACM, New York, NY, USA, 12 pages. https://doi.org/10.1145/3589334.3645328."
\end{textblock}

\maketitle

%% paper structure %% COMMENT THIS WHEN FINALIZED
% \setcounter{section}{-1}
% \section{Structure}
% \begin{enumerate}

% \item Last-mile performance over time [Andre]
%     \begin{enumerate}
%     \item active measurements we have been running in Edinburgh and Munich dishys (iRTT + traceroute). (i) time-series correlation, (ii) correlation between multiple dishes simultaneously (iii) 
%     \item Possible to correlate IRTT, packet losses, and satellite movements similar to~\cite[Fig. 7]{kassem2022}?
%     % \item Real-time communication: video / voice. How would a test setup look like?
%     \item Impact on last-mile. TCP/QUIC workload?
%     \end{enumerate}

% \item Impact of mobility [Hendrik]
%     \begin{enumerate}
%     \item mobility Starlink vs cellular?
%     \end{enumerate}

% \end{enumerate}

%% Import sections
% \vspace*{-1em}

\section{Introduction}

Over the past two decades, the Internet's reach has grown rapidly, driven by innovations and investments in wireless access~\cite{5g_look, mobiwac-mmwave, variegated-5g} (both cellular and WiFi) and fiber backhaul deployment that has interconnected the globe~\cite{arnold2020much, yuchen2019sigcomm, ching2015arewe, yeganeh2020first, dang2021}.
%
% Recent research explorations are further improving the operations of these (terrestrial) technologies by utilizing high-frequency wireless spectrum that boosts significantly higher data rates even during mobility~\cite{variegated-5g} and by deploying fiber backhaul to the edge of the network~\cite{dang2021}.
%
Yet, the emergence of Low-Earth Orbit (LEO) satellite networking, spearheaded by ventures like Starlink~\cite{starlink}, OneWeb~\cite{oneweb}, and Kuiper~\cite{kuiper}, is poised to revolutionize global connectivity.
LEO networks consist of mega-constellations with thousands of satellites orbiting at 300--2000~km altitudes, promising ubiquitous \textit{low latency} coverage worldwide.
Consequently, these networks are morphing into ``global ISPs'' capable of challenging existing Internet monopolies~\cite{starlink2023d2c}, bridging connectivity gaps in remote regions~\cite{ma2022network, starlink-maritime}, and providing support in disaster-struck regions with impaired terrestrial infrastructure~\cite{starlink-ukraine}.
%
% Such mega-constellations may feature thousands of satellites deployed $\leq$2000~km above Earth's surface traveling at $\approx$~27000~km/h in orbit to offer global \textit{low-latency} Internet access.

\emph{Starlink} from SpaceX stands out with its expansive fleet of 4000 satellites catering to 2M+ subscribers across 63 countries~\cite{starlink-subs, starlink-wiki}.
% The most prominent and popular Internet provider using LEO mega-constellations is \emph{Starlink} from SpaceX.
The LEO operator plans to further amplify its coverage and quality of service (QoS) by launching $\approx$ 42,000 additional satellites in the coming years~\cite{businessinsider2023starlink}.
%
% The first satellites for the network were launched in 2018 and have more than 3000 satellites operational (as of May 2023) with plans to launch some 40,000 more over the next few years (cf. \cref{sec:relatedwork}) and serve 1.5M+ subscribers worldwide~\cite{starlink-subs}.
% \textcolor{red}{JO: The sentence suggests that serving 1.5M+ subscribers is \textit{planned} but the tweet suggests that this number holds today.}
% \textcolor{red}{[Details about the Starlink network here]}.
%
% Its significant global reach and impressive connectivity have made Starlink a viable alternative to the existing terrestrial Internet infrastructure.
% Additionally, the network has found unique utility in access to remote previously-unconnected regions~\cite{ma2022network}, maritime communications~\cite{starlink-maritime}, circumventing local governmental restrictions~\cite{starlink-ukraine}, and more. 
However, despite significant global interest and the potential to impact the existing Internet ecosystem, only limited explorations have been made within the research community to understand Starlink's performance.
%
% The task itself is challenging as accurately ascertaining the LEO network's performance necessitates vantage points across the entire globe, as connectivity can be affected by factors unique to each region (e.g. orbital coverage, ground infrastructure density, etc.).
The challenge stems from a lack of global vantage points required to accurately gauge the network's performance since factors such as orbital coverage, density of ground infrastructure, etc., can impact connectivity across regions.
Initial studies have resorted to measurements from a handful of geographical locations~\cite{kassem2022,michel2022,ma2022network,lopez2022} or extrapolated global performance through simulations~\cite{hypatia} and emulations~\cite{lai2023starrynet}.
However, the community agrees on the limited scope of such studies and has made open calls to establish a global LEO measurement testbed to address this challenge~\cite{satnetlab, pan2023measuring, tanveer2023making}.
Some researchers have navigated around this hurdle by exploring alternative measurement methods, e.g., by targeting exposed services behind user terminals~\cite{izhikevich2023democratizing}, by mining measurements on social media platforms~\cite{taneja2023viewing}, or by recruiting users in select regions~\cite{dissectingleo}.
While innovative, we argue that these techniques are insufficient to uncover the intricacies affecting the network, specifically its capability to support applications.  

This paper addresses this knowledge gap and provides the first comprehensive multi-faceted measurement study on Starlink.
Our work is distinct from previous research in several ways.
Firstly, 
% unlike previous investigations that relied on vantage points in a small set of locations, 
we examine the global evolution of the network since 2021 by analyzing the M-Lab speed test measurements~\cite{mlab-pub} from 34 countries (largest so far).
We complement our investigation 
% the coarse-grained M-Lab measurements 
through active measurements over 98 RIPE Atlas~\cite{ripe-atlas} probes in 21 countries and 
% from Starlink vantage points over RIPE Atlas and our own fully controlled environment.  % Our analysis relies on more than 12M measurements obtained over several months through diverse sources, as outlined above and elaborated in \Cref{sec:methodology}.
%
% Additionally, we 
conduct high-resolution experiments over controlled terminals in two European countries to investigate real-time web application performance and factors impacting Starlink's last-mile access.
%
% We shed light on several open questions surrounding Starlink's last-mile ``bent-pipe''~\cite{kassem2022} and its impact on the performance of Internet-supported latency-critical applications.
% We thus offer the first study to provide a deep understanding of the Starlink network as we answer several questions, such as \emph{ques 1}, \emph{ques 2}, and \emph{ques 3}.
% As such, this paper contributes to a deeper understanding of the Starlink network in the following ways:
Specifically, we make the following contributions in this work.
%
%\smallskip \noindent

\smallskip
\noindent
\textbf{(1)} We present a longitudinal study of global Starlink latency and throughput performance from M-Lab users
% from the perspective of M-Lab users globally  (
in \Cref{sec:global}.
Our analysis, incorporating $\approx$ 19.2~M samples, reveals that Starlink performs competitively to 
% latency and goodput are at par with 
terrestrial cellular networks.
However, its performance varies globally due to infrastructure deployment differences, and is dependent on
% and considering that the LEO network's performance correlates strongly with 
the density and closeness of ground stations and Points-of-Presence (PoPs).
%\smallskip \noindent
We also observe signs of \emph{bufferbloating} as Starlink's latency significantly increases under traffic load.

\smallskip
\noindent
\textbf{(2)} We assess and compare the 
% impact of increased latency under traffic load and latency changes at 15s intervals with Starlink on the
performance of real-time web applications, specifically Zoom video conferencing and Amazon Luna cloud gaming, to terrestrial networks (\Cref{sec:cloud-gaming}).
%
% We set up an automated system running gaming workloads over Amazon Luna~\cite{luna} and conduct several runs over Starlink, terrestrial fixed-line access, and cellular (5G) network in the region. 
%
We find that, under optimal conditions, Starlink is capable of supporting such applications, matching the performance over cellular; however, we do observe some artifacts due to the network's periodic reconfigurations.
% perform for a demanding cloud-based gaming application.

\smallskip
\noindent
\textbf{(3)} We perform targeted measurements from Starlink RIPE Atlas~\cite{ripe-atlas} probes and leverage their diverse locations to characterize the satellite last-mile performance (\Cref{subsec:ripeatlas}).
We find that the ``bent-pipe'' (terminal $\leftrightarrow$ satellte $\leftrightarrow$ ground station) latency within the dense 53$\degree$ shell remains consistent worldwide ($\approx$ 40~ms), and is significantly lower to yet incomplete 70$\degree$ and 97.6$\degree$ orbits.
%
% However, the performance in regions lying beyond these shells and served by, as yet incomplete 70$\degree$ and 97.6$\degree$ shells are significantly lower, with intermittent connectivity.
We also find evidence of Starlink inter-satellite links (ISLs) delivering superior performance to terrestrial paths for connecting remote regions. 

\smallskip 
\noindent
\textbf{(4)} Our fine-grained measurements from terminals in two European countries 
% to uncover the patterns and impact of Starlink's periodic reconfiguration intervals.
%
% We 
confirm that Starlink performs network reconfigurations every 15~s, leading to noticeable latency and throughput degradations at sub-second granularity.
%
% One connects to the 53$\degree$ shell while the other one can be shielded to confine its communication to the 70$\degree$ and 97.6$\degree$ orbits (\Cref{subsec:controlled}).
%
% Our fine-grained probing using \texttt{irtt} and \texttt{iperf} reveals that Starlink follows strict 15s ``reconfiguration intervals'' and we observe short-lived throughput drops of several orders at interval boundaries.
%
%
By correlating data from our terminals in Germany (within 53$\degree$) and Scotland (restricted to 70$\degree$ and 97.6$\degree$ coverage), we find that the reconfigurations are globally synchronized and likely independent of satellite handovers.
% argue that these intervals are used by Starlink satellites for managing connectivity with the client terminals.

% paper veto gain insights into some operational aspects of the Starlink network and dissect Starlink's last-mile performance.
%

% For (3) and (4), we also compare Starlink performance to other terrestrial network alternatives (cellular and fixed line access).

% \textcolor{red}{JO: Check the above and add selected findings below.}
% Through our measurement analysis study in this paper, we come up with the following new and significant findings:
% \begin{itemize}
% \item Starlink's performance correlates strongly with the density of supporting ground infrastructure close to its users; due to lacking deployments, its performance varies globally (\Cref{sec:global})
% \item The ``bent-pipe'' latency of Starlink's 53$\degree$ shell is similar around the world; the difference to the, as yet, incomplete 70$\degree$ and 73$\degree$ shells however stark (\Cref{subsec:ripeatlas})
% \item Ditto for section 5.2 (\Cref{subsec:controlled})
% \item Under best-case conditions, Starlink is already capable of supporting real-time applications on par with the latest cellular standards (\Cref{sec:cloud-gaming})
% \end{itemize}

% \smallskip
%Our paper is organized as follows.
% In the next section, we give an overview of the Starlink network by way of background and discuss prior measurement based studies focusing on Starlink.
Leveraging multi-dimensional, global, and controlled high resolution measurements, our findings distinctively advance the state-of-the-art by illuminating Starlink's performance and the influence of internal network operations on real-time applications.
To foster reproducibility and enable future research, we publish our $>$ 300~GB collected dataset and associated scripts at \cite{paper-code} and \cite{paper-dataset}.

\section{Background} \label{sec:background}

% In this section, we provide an overview of the Starlink network, its anatomy, and the key components involved. Additionally, we review previous research efforts that have focused on analyzing the Starlink network constellation.

% Introduction to the section
\begin{figure}[t!]
    \centering
    \includegraphics[width=0.9\columnwidth]{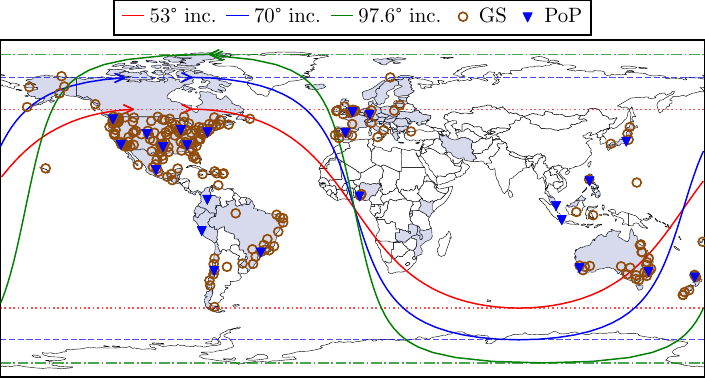}
    \vspace*{-1em}
    \caption{\label{fig:mlab_shells} Orbits of three Starlink inclinations and crowdsourced Ground Station (GS) and Point-of-Presence (PoP) locations~\cite{starlink-gs-pop-unofficial}. Shaded regions depict Starlink's service area.}
    \vspace*{-1.5em}
\end{figure}

% \subsection{Starlink Network Overview} \label{subsec:background}
Starlink is a LEO satellite network operated by SpaceX that aims to provide global Internet coverage through a fleet of satellites flying at $\approx$~500~km above the Earth's surface.
%
% Starlink has launched over 4,000 satellites in orbital shells of different inclination angles and multiple planes over different generations to date, with plans to expand up to 42,000 satellites in the future~\cite{starlink-wiki}.
%
The majority of Starlink's operational 4000 satellites lie within the 53$\degree$ shell, which only covers parts of the globe (see \Cref{fig:mlab_shells}).
The 70$\degree$ and 97.6$\degree$ orbits allow serving regions near the poles. %however, these only have a limited number of satellites
These other shells however have fewer satellites
%
% For more details, 
(see \Cref{app:orbits}, \Cref{tbl:starlink_orbits} for constellation details).
%
%
% As of May 2023, Starlink has launched v0.9, v1.0, v1.5, and a few v2-mini (second generation \cite{starlink-v2-mini}) satellites.
% in which the latter belong to the second generation of Starlink \cite{starlink-v2-mini}. 
% In an iterative fashion, the v1.0 satellites where being manufactured and launched while the design for the v1.5 is being worked on until mid 2021 when its deployment started. 
% The first shell of 53° is populated with v1.0 satellites while rest are v1.5. 
% The major difference between the two versions is that v1.5 features laser inter-satellite link capabilities, although its wide-scale activation is still in process.
%
%

\begin{figure}[t!]
    \centering
    \includegraphics[width=0.42\textwidth]{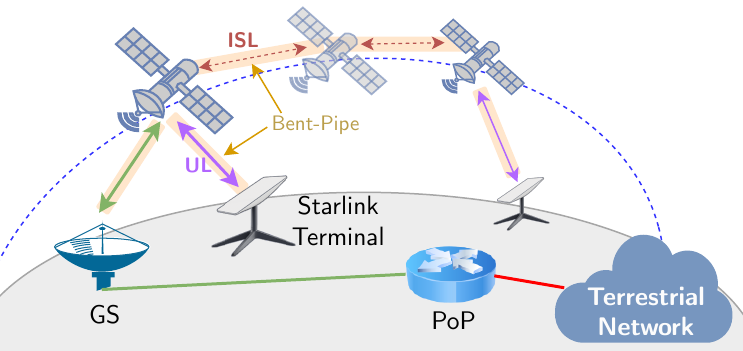}
    \vspace*{-1em}
    \caption{\label{fig:starlink-bentpipe}
      %Illustration of connectivity with Starlink.
      Starlink follows ``bent-pipe'' connectivity as traffic traverses the client-side terminal, one or more satellites via inter-sat links (ISLs), nearest ground station (GS), ingressing with the terrestrial Internet via a point-of-presence (PoP).}
      \vspace*{-2em}
\end{figure}

\Cref{fig:starlink-bentpipe} shows the cross-section of Starlink end-to-end connectivity.
To access the Internet over the Starlink network, end-users require a dish, a.k.a. ``Dishy''\footnote{We use ``Dishy'' and ``user terminal'' interchangeably in the paper.}, that communicates with satellites visible above 25$\degree$ of elevation through phased-array antennas using Ku-band (shown as User Link (UL)).
Starlink satellites, equipped with multiple antennas subdivided into beams, can connect to multiple terminals simultaneously~\cite{iyer2022system} and relay all connections to a ground station (GS) on a Ka-band link (shown in green).
%
% The link between the Starlink terminal and the satellite operates .
% The satellite forwards the data to the nearest GS on a Ka-band link.
The connection forms a direct ``bent-pipe'' in case the terminal and GS lie within a single satellite's coverage cone; otherwise, the satellites can relay within space to reach far-off GSs via laser inter-satellite links (ISLs), forming an ``extended bent-pipe''.
% Both the dish and GS must be within the satellite's coverage cone to let them interconnect.
%
Note that not all Starlink satellites are ISL-capable and it is difficult to effectively estimate ISL usage as Starlink satellites have no user visibility at IP layer and, therefore, do not show up in \texttt{traceroute}s~\cite{pan2023measuring}.
%
% In this work, we highlight in a case study investigating Reunion Island (\textcolor{red}{sec xx}) the use of ISLs to connect ``terrestrially inaccessible'' regions.
%

Finally, the GSs terrestrially relay traffic from satellites to Starlink points-of-presence (PoPs), which route it to the destination server via terrestrial Internet~\cite{starlink-pop2021,pan2023measuring}.
The public availability of GS deployment information differs across countries.
% Accessibility to information regarding GSes largely depends on the regulatory bodies of the local governments.
No official source exists, so we rely on crowdsourced data for the geolocations of GSs and PoPs~\cite{starlink-gs-pop-unofficial}, which is also shown in \Cref{fig:mlab_shells}.
% The satellite connection from the terminal to the PoP (the green path in \Cref{fig:starlink-bentpipe}) is referred to as the ``bent-pipe''.
%
% Little is known, as of now, on the operations of this ``bent-pipe'' and its impact on end-to-end performance.
% In other words, the bent-pipe is effectively a black-box.
% Our intention in this paper is to infer its behavior through diverse type of measurements.
% Within this paper, we shed light on its characteristics using diverse approaches.

% The Starlink network consists of the satellite constellation, terminals, ground stations (GSes), and Points-of-Presence (PoPs).
% \Cref{fig:mlab_shells} provides a combined view of the groundtrack of different orbital inclinations and the geographical distribution of Starlink GSes and PoPs worldwide.

% For our work, we see these GS and PoP numbers as minimum reference. \textcolor{red}{PR: please check if the way I described this is appropriate}

% Omit the "operational"
% Remove #Sats
% \textit{Satellite in position} means a satellite which has reached it's filed orbital altitude and position in its orbital plane. However, it is important to keep in mind that the 'operational orbit' of a satellite and whether or not a satellite is in operational service might be independent to one another.}

\section{Measurement Methodology} \label{sec:methodology}

\subsection{Global Measurements} \label{sec:methodology:global}
% \subsection{Measurement Lab (M-Lab)} \label{sec:methodology:mlab}
% \vspace{-0.25cm}

\begin{figure}[t]
	\includegraphics[width=0.88\columnwidth]{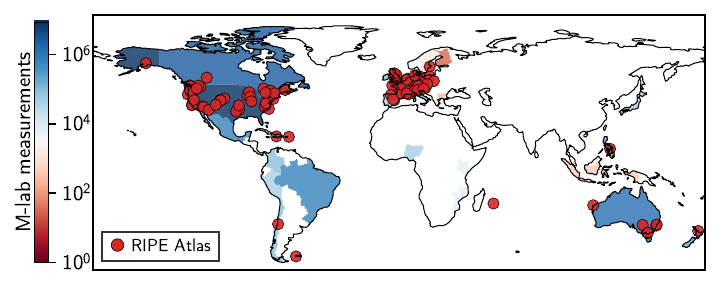}
        \vspace*{-1.5em}
	\caption{\label{fig:mlab_probecities} Overview of global Starlink measurements in this study. Heatmap denotes M-Lab speedtest measurement densities. Starlink RIPE Atlas probes are shown as red circles.}
 \vspace*{-2em}
\end{figure}

\noindent\textbf{\emph{Measurement Lab (M-Lab)}}
M-Lab~\cite{mlab-pub} is an open-source project that allows users to perform end-to-end throughput and latency speed tests from their devices to 
% M-Lab servers.
%
% The platform offers 
500+ servers in 60+ metropolitan areas~\cite{mlab-platform}.
% hosted in ``off-net'' data centers or ISP-cloud peering locations~\cite{mlab-platform}.
% and
%
% With this setup, the measurements over M-Lab 
% therefore provides a complete user-to-content server perspective.
%
% Since the initiative is backed by Google, the 
%
% Google performs M-Lab throughput measurement when a user searches for ``speedtest'' keyword, which is also the primary source of initiations.
Google offers M-Lab measurements when a user searches for "speed test"~\cite{mlab-google}, serving as the primary source of measurement initiations~\cite{mlab-ukraine, activeMeas-imc2022, mlab-pub}.
% The M-Lab client is integrated with Google search~\cite{mlab-google}, which is also the primary source of measurement initiations~\cite{mlab-ukraine, activeMeas-imc2022, mlab-pub}.
% and the servers are present in all Google data centers.
%
At its core, M-Lab uses the Network Diagnostic Tool (NDT)~\cite{mlab-ndt}, which measures uplink and downlink performance using a single 10~s WebSocket TCP connection.
The platform also records fine-grained transport-level metrics (\texttt{tcp\_info}), including goodput, round-trip time (RTT) and losses, 
% and other transport-level metrics.
% 
% Additionally, the tool reports the 
along with IP, Autonomous System Number (ASN), and geolocation of both the end-user device and the selected M-Lab server.
%
% This localization method is not able to accurately detect a Starlink user's location and reports, in many cases, only one city per country.
% We approach our analysis with this fact in mind; see \Cref{sec:discussion} for further discussion on this issue.
%
% \footnote{M-Lab uses IP geolocation, which is known for yielding potentially inconsistent results; see \Cref{sec:discussion}.}.
%
% In this work, we use the M-Lab network measurements triggered by Starlink users from August 2021 to April 2023, totalling 20 months.
% \Cref{fig:mlab_probes_overtime} plots the Starlink launches and thus the network growth and how the M-Lab probe count considered in this paper evolved over time.
%
% We identify the measurements made by Starlink users through client ASN (AS14593). 
We identify measurements from the Starlink clients via their ASN (AS14593).
%
% Before early 2021, Starlink used Google ASN (AS36492) for its operations which makes Starlink users indistinguishable from Google cloud servers in M-Lab measurements~\cite{starlink-internet-infra,kassem2022}. 
% %
% We avoid the potential effect of this transition by using June 2021 as the measurement start point for our analysis.
%
The M-Lab dataset includes samples from 59 out of 63 countries where Starlink is operational.
%
% However, some countries where Starlink became active at the beginning of 2023, e.g., Haiti, Ecuador, and Finland, only had a few reported measurements.
%
We restrict our analysis to \texttt{ndt7} measurements, which use TCP BBR
% and remove all data points gathered with earlier versions, such as \texttt{ndt5}, which used TCP Reno/Cubic.
%
% We avoid such vagrant data points impact the accuracy of our analysis, 
% To maintain statistical significance, we include only 
and countries with \emph{at least 1000 measurements} since June 2021 (launch of Starlink \texttt{v1.0} satellites~\cite{starlink-v1.0}), resulting in 19.2~M M-Lab measurement samples from 34 countries.
M-Lab infers the approximate location of Starlink users from their public IP which is assigned by the PoP~\cite{starlink-geoip-csv}.
As a result, all speed tests across countries are mapped to a city, except for USA and Canada, which are sub-divided into multiple regions.
% }
%
% While this is a reasonable approach for most terrestrial networks, it may not be the case for Starlink, as the assigned PoP may be situated far from the client's location, thereby including additional terrestrial path lengths in the measurement.
% Note that this may include inflated latencies when analyzing Starlink, as the assigned PoP may be situated far from the client's location, thereby including additional terrestrial path lengths in the measurement.
%
While we examine such artifacts by contrasting the M-Lab and RIPE Atlas results (\Cref{subsec:ripeatlas}),
we approached our analysis with caution, particularly when examining fine-grained region-specific insights.
\headline{RIPE Atlas}
% While M-Lab offers broad coverage, it does not provide visibility into the Starlink bent-pipe operation and performance.
% %
% The M-Lab NDT test suite includes \texttt{traceroute}, but it is triggered from the server-side and only resolves until the PoP, before the bent-pipe.
% %
% We fill this gap by conducting active measurements over the RIPE Atlas, which is
% \footnote{While our controlled experiments (see \Cref{sec:methodology:controlled}) allow us to investigate the last-mile performance, it is limited to two locations and, therefore, do not provide a global perspective.}.
%
RIPE Atlas is a measurement platform that the networking research community commonly employs for conducting measurements~\cite{ripe-atlas}.
% RIPE Atlas is a global Internet measurement network and is prominently used within the networking research community for conducting measurements.
%
The platform comprises thousands of hardware and software probes scattered globally, enabling users to carry out active network measurements such as \texttt{ping} and \texttt{traceroute} to their chosen endpoints.
% includes thousands of small hardware and software probes connected to the Internet globally.
% all over the world.
%
% Users can perform active network measurements (most notably \texttt{ping}, \texttt{traceroute}, DNS resolution, etc.) from these probes to endpoints of their choice.
%
%As of May 2022, the platform includes 10K+ probes in $\approx$ 2.5K ASes available in 200+ countries.
%
% 98 Starlink probes hosted in RIPE Atlas across 21 countries have been used for our measurements.
We utilized 98 Starlink RIPE Atlas probes across 21 countries (see \Cref{fig:mlab_probecities}).
%
% Figure \ref{fig:mlab_probecities} shows the geo-distribution of these probes.
%
%The majority of the probes are located in the US (35) and Europe (22), but we also find coverage in Oceania and South America.
%
% Overall, our active measurements over Starlink-enabled RIPE Atlas VPs provide us last-mile visibility in over 14 countries -- augmenting our M-Lab analysis.
%
%\smallskip
%\noindent \textbf{Experiments.} 
%
% Our RIPE experiments serve two objectives:
% \emph{(a)} investigate the satellite bent-pipe performance and \emph{(b)} understand the routing and peering behavior of Starlink with the Internet.
%
% As such, w
% As targets for our measurements from RIPE Atlas probes,
Our measurement targets were 145 data centers from \emph{seven} major cloud providers -- Amazon EC2, Google, Microsoft, Digital Ocean, Alibaba, Amazon Lightsail, and Oracle (see \Cref{app:cloud-dcs}).
The chosen operators represent the global cloud market
%
% The chosen operators are widely used, globally deployed and offer a well established network infrastructure.
%
~\cite{arnold2020much,yuchen2019sigcomm,li2010cloudcmp,yeganeh2020first}
% and have shown to be highly pervasive and reachable, thanks to their extensive investments in WAN deployments and ISP peerings~\cite{corneoRIPE,li2010cloudcmp,yeganeh2020first}.
%
% Recent studies have shown Starlink PoPs to be co-located with cloud data centers~\cite{kassem2022} and at Internet eXchange Point (IXP) facilities~\cite{izhikevich2023democratizing}. 
% we find PoPs in locations where Google does not have a data center presence (e.g., Africa in \Cref{fig:mlab_shells}).
%
% We verify this claim in our initial investigations by resolving the ASN of the first IP address in our \texttt{traceroute}, and found it to be the case [\textcolor{red}{@Rohan: Needs verification}].
%
% To accomodate for this, we included all available Google data centers as endpoints in our measurements.
%
% Since Starlink has reportedly PoPs available in locations where Google does not have a data center offering (most notably in Africa),
% Due to the existing peering agreements between cloud providers and IXPs~\cite{ching2015arewe,corneoRIPE}, our diverse cloud provider selection 
and ensure that our endpoints are close to Starlink PoPs, which are usually co-located with Internet eXchange Point (IXP) or data center facilities~\cite{izhikevich2023democratizing, kassem2022}.
We perform ICMP \texttt{traceroute}s from Atlas probes to endpoints situated on the same or neighboring continent.
We extract and track per-hop latencies between Starlink probe terminal-to-GS (identified by static \texttt{100.64.0.1} address), GS-to-PoP \texttt{(172.16/12}) and PoP-to-endpoint at 2~s intervals~\cite{pan2023measuring}.
%
% Regardless of the targeted endpoint, all \texttt{traceroute} measurements report the latency of the bent-pipe -- allowing us to track its behavior at roughly 2s intervals.
%
% Depending on the location, a probe ran between 40 to 50 \texttt{traceroute}s to all relevant endpoints every hour -- totaling $\approx$ 500 per day.
%
% We correlate end-to-end latency correctness through ICMP \texttt{ping} measurements at similar intervals.
% which allow us to correlate latencies with measurements from M-Lab dataset from the same location as the Atlas probes.
% 
Additionally, we also extract semantic location embeddings in reverse DNS PTR entry, e.g. \texttt{tata-level3-\underline{seattle}2.level3.net} to further improve geolocation accuracy~\cite{conext2021learning}. 
%reverse DNS PTR for each of the RIPE Atlas probes used in our measurements, we extract the first IP address outside the Starlink network (ASN 14593), and 
% Specifically, we perform a reverse DNS lookup on PoP IP address. Often a domain name may consist of a geo-location identifier, for e.g. \emph{tata-level3-seattle2.level3.net}. If the RTT difference between this IP and the POP IP <= 1ms, we can safely assume these to be co-located.
%More details on the measurements for RIPE Atlas can be found in \Cref{tbl:meas-overview}
Our experiments over \emph{ten} months (Dec 2022 to Sept 2023) resulted in $\approx$ 1.8~M measurement samples.

\begin{comment}
\begin{figure}[t]
  \centering
  \includegraphics[width=\columnwidth]{figures/gaming.drawio.pdf}
  \vspace{-0.8cm}
  \caption{\label{fig:gaming-architecture} %
  We utilize Amazon Luna~\cite{luna} as the cloud gaming platform and use an instrumented game client to record video stream information.
  We measure performance over a wired 1~Gbps network, 5G, and Starlink.}
  \vspace{-0.5cm}
\end{figure}
\end{comment}

\vspace*{-1em}
%subsubsection*
\subsection{Real-time Web Application Measurements} \label{sec:methodology:applications}

% Previously, we have looked at the performance of Starlink at a global level and scrutinized the bent-pipe to get an insight into its characteristics.
%
% We conclude this work by challenging Starlink with an application that demands low-latency and constant-throughput data delivery to explore its current performance boundaries and potentials.

\noindent\textbf{\emph{Zoom Video Conferencing.}}\label{sec:methodology:zoom}
% We experimented with Zoom video conferencing~\cite{zoom} due to its popularity on the Internet~\cite{feldmann2020zoom} as well as its latency and bandwidth-critical operational requirements.
% Video conferencing is a popular and demanding Internet application.
% It has seen a surge of usage in recent years~\cite{feldmann2020zoom}
% and requires a low end-to-end delay to enable natural conversations.
% %
% We measured the performance of the popular video conferencing application Zoom~\cite{zoom} by setting up a call with two parties.
We set up a Zoom call between two parties, one using a server with access to an unobstructed Starlink dish and high-speed terrestrial fiber over 1~Gbps Ethernet.
The other end was deployed on an AWS machine located close to the assigned Starlink PoP.
% First, Zoom was run on an university server that had access to an unobstructed Starlink dish and to a high-speed Internet connection over an 1~Gbps Ethernet link.
% The second Zoom instance was run on an AWS instance hosted close to the Starlink PoP assigned to our terminal ($\approx$~1~ms RTT).
% We configured a virtual camera and microphone on both machines that were fed by a pre-recorded video showing a single person talking to conduct duplex tests.
We set up virtual cameras and microphones on both machines, which were fed by a pre-recorded video of a person talking, resulting in bidirectional transmission.
Both machines were time-synchronized to local stratum-1 NTP servers and
% within their local networks.
we recorded (and analyzed) Zoom QoS leveraging the open-source toolchain from \cite{michel2022zoom} that provides sub-second metrics.
% for analysis that yield frame and packet-level metrics.

\headline{Cloud Gaming}\label{sec:methodology:gaming}
%
%
% For this task, w
% Cloud gaming enables users to play high-quality games on low-end devices by offloading the game and rendering logic to a remote server.
%
We also experiment with cloud gaming due to its demanding high throughput and low delay requirements~\cite{pruning-hotnets2020}. 
% The application is the most demanding amongst realtime multimedia services~\cite{pruning-hotnets2020} as it requires consistently high throughput and low delay to maintain an enjoyable user experience.
%
% Interruptions are immediately noticed by users and negatively affect their experience.
%
We leverage the automated system by \citet{iqbal2021} to evaluate the performance of playing the racing game ``The Crew'' on the Amazon Luna~\cite{luna} platform.
%
% \Cref{fig:gaming-architecture} shows our setup.
%
Specifically, we measure using a customized streaming client that records end-to-end information about media streams, such as frame and bitrate.
The system utilizes a bot that executes in-game actions at pre-defined intervals that trigger a predictable and immediate visual response.
In post-processing, the analysis system computes the \emph{game delay} as the time passed since the input action was triggered.
%
% We set up their system to evaluate the performance of playing the racing game ``The Crew'' on the Amazon Luna~\cite{luna} platform.
% In our case, the bot repeatedly accelerated the car and engaged the brakes, triggering the car's brake lights.
% Amazon Luna serves games at a resolution of up to 1920$\times$1080 (full-HD) at 60 FPS but adaptively reduces the resolution and bitrate to, e.g., 1280$\times$720.
% We set up their system to evaluate .
Amazon Luna serves games at 60~FPS and 1920$\times$1080 resolution which adaptively reduces to 1280$\times$720.
%
%  depicts our evaluation setup and the measured contributors to the overall game delay.
We ran the game streaming client on the same machine as the Zoom measurements, additionally setting up a 5G modem to compare Starlink against cellular.
%We ran the game streaming client on a desktop PC attached to a 60~FPS full-HD monitor and compared performance of three different networks:
%A high-speed 1~Gbps Ethernet (``terrestrial'' link), a 5G modem connected to the network of a large national network operator,
% and an Ethernet connection to an unobstructed Starlink dish. %
%
% Our client machine was located in Europe, and the Luna game server was in an AWS data center close to the Starlink PoP assigned to our terminal ($\approx$ 1~ms RTT).
Similar to Zoom, the Luna game server was on AWS server close to our Starlink PoP ($\approx$ 1~ms RTT).

\vspace*{-1em}
\subsection{Targeted Measurements} \label{sec:methodology:controlled}

\begin{figure}[t]
    \centering
        \includegraphics[width=0.6\columnwidth]{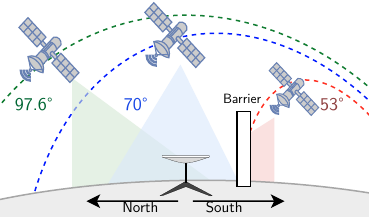}
    \vspace*{-1em}
    \caption{\label{fig:last-mile-fov-schematic} Field-of-view experiment setup. Dishy, deployed at a high latitude location, is obstructed by a metal shielding, which restricts its connectivity to the 70$\degree$ and 97.6$\degree$ orbits.}
    \vspace*{-2em}
\end{figure}

% \headline{Controlled Experiments}
%
A significant limitation of RIPE Atlas measurements is their lack of sub-second visibility, which is 
% Analyzing the network performance at such level is 
essential for understanding the intricacies of Starlink network.
% , particularly the bent-pipe, and its impact on the application layer performance.
% To allow us to obtain microscopic understanding, 
To combat this, we orchestrated a set of precise, tailored, and controlled experiments, utilizing two Starlink terminals as vantage points (VPs) situated in Germany and Scotland. 
Our dish in Germany connects to the 53$\degree$ shell while Scotland dish, due to the high latitude location, can be shielded to confine its communication to the 70$\degree$ and 97.6$\degree$ orbits.
We do this by placing a metal sheeting
% \footnote{Metal sheeting was chosen due to its ability to act as a Faraday shield, blocking the RF emissions from satellites.} 
as Faraday shield barrier at the South-facing angle of the terminal, which obstructs its view from the 53$\degree$ inclinations (see \Cref{fig:last-mile-fov-schematic}).
% This obstructed the 53$\degree$ inclinations.
%
We verify with external satellite trackers~\cite{celestrak2023, starlinksx2023} that the terminal only received connectivity from satellites in 97$\degree$ or 70$\degree$ inclinations, which resulted in brief \textit{connectivity windows} followed by periods of no service.
%
% With far fewer (see Table \ref{tbl:starlink_orbits}) satellites in the unobstructed inclinations, this results in large periods of no service, interrupted by brief \textit{connectivity windows} when satellites in the 97$\degree$ or 70$\degree$ inclinations were passing overhead. 
% \noindent \textbf{Measurement Tools.}
% The two primary measurement tools used for the controlled experiments were 
We performed experiments using the
Isochronous Round-Trip Tester (\texttt{irtt})~\cite{irtt} and \texttt{iperf}~\cite{iperf} tools. 
The \texttt{irtt} setup records RTTs at high resolutions (3~ms interval) by transmitting small UDP packets.
% between the client and server. 
%
% We set the inter-packet delay to 3~ms which achieves finest possible granularity without causing congestion. 
%
The \texttt{irtt} servers were deployed on cloud VMs in close proximity to the assigned Starlink PoP of both VPs (within 1~ms) --  minimizing the influence of terrestrial path on our measurements. 
%
% We observed consistent \texttt{ping} RTTs of less than 1~ms between the \texttt{irtt} server and the PoP.
%
We used \texttt{iperf} to measure both uplink and downlink throughput and record performance at 100~ms granularity.
% of the Starlink connection. 
%
Simultaneously, we polled the gRPC service on each terminal~\cite{starlink-grpc-tools} every second to obtain the connection status information.
% We set the \texttt{iperf} measurement interval to the minimum allowed 100~ms and measure
%  we were able to gain insight into the sub-second performance of the connection. We measured 
% the true bandwidth of Starlink end-to-end connection.
  % and measuring the traffic arriving at the receiving end.

% \noindent \textbf{Restricted Field-of-View Experiments.}

\begin{figure*}[!t]
	\begin{minipage}{0.65\linewidth}
		\begin{subfigure}[htbp]{0.495\textwidth}
			\includegraphics[width=\textwidth]{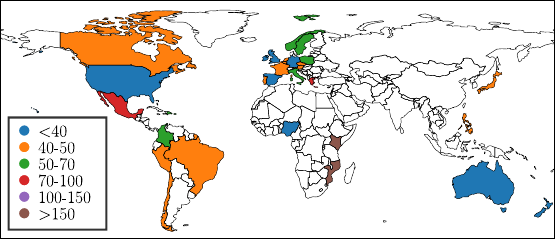}
			% \caption{Starlink}
		\end{subfigure}
		\hfill
		\begin{subfigure}[htbp]{0.495\textwidth}
			\includegraphics[width=\textwidth]{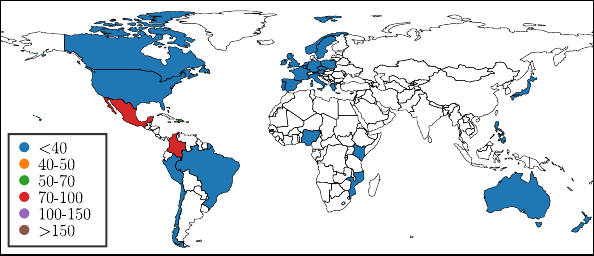}
			% \caption{Top-3 serving ISPs}
		\end{subfigure}
		% \begin{subfigure}[htbp]{0.4\textwidth}
		% 	\includegraphics[width=\textwidth]{figures/mlab/state_2023-10-01/mlab_mnocell-filtered_mapped_medianminrtt_oneyear_heatmap.pdf}
		% 	\caption{Heatmap of median MinRTTs over one year period of all cellular ISPs filtered for mobile devices\textcolor{red}{todo: write text}}
		% \end{subfigure}
		\vspace*{-1em}
		\caption{\label{fig:mlab_heatmaps} Median of minimum RTT (in ms) of devices connected via Starlink (left) and top-3 serving ISPs (right) in the same country to the nearest M-Lab server. }
	\end{minipage}%
	\hfill
	\begin{minipage}{0.34\linewidth}
		\centering
		\includegraphics[width=\columnwidth]{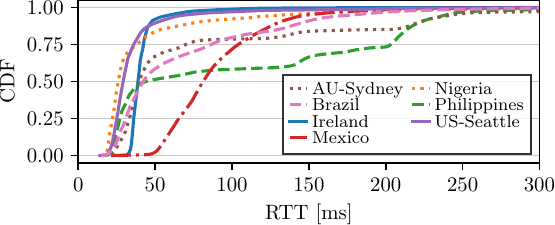}
		\vspace*{-2.3em}
		\caption{\label{fig:mlab_cdf_minrtt_glob} Starlink latency distribution from select cities in each continent.}
	\end{minipage}
	\vspace{-1em}
\end{figure*}

% \textcolor{red}{
% We hypothesized that a deeper insight into the Starlink last-mile could be achieved by observing the fine-grained behavior of a Dishy in a high latitudinal location, where only the 70$\degree$ and 97.6$\degree$ inclinations would be visible. With only access to a Dishy at a 56$\degree$ latitude (which still connects to the 53$\degree$ inclinations), we  placed a barrier created from metal sheeting%
% \footnote{Metal sheeting was chosen due to its ability to act as a Faraday Shield, blocking the RF emissions from satellites.}
% around it at a South-facing angle. This obstructed the 53$\degree$ inclinations.
% %
% With far fewer (see Table \ref{tbl:starlink_orbits}) satellites in the unobstructed inclinations, this results in large periods of no service, interrupted by brief \textit{connectivity windows} when satellites in the 97$\degree$ or 70$\degree$ inclinations were passing overhead. To obtain data on these connectivity windows, we polled the gRPC service that is present on the Dishy \cite{starlink-grpc-tools} at a rate of once per second. Among the other status information reported through this service, a ``state'' field indicates when the dish is online.
% }

% \noindent \textbf{Co-Located Dish Experiments.}

\section{Global Starlink Performance} \label{sec:global}

We use the minimum RTT (minRTT) reported during \texttt{ndt7} tests to the closest M-Lab server globally as baseline network performance.
Note that minRTT is not affected by queuing delays due to bandwidth-capped data transfers during speed tests.
%  which results in elevated latencies.
%
% We compare the Starlink LEO network to a wireless mobile network (i.e., WiFi or cellular) by comparing it against the terrestrial serving-ISP infrastructure in that country.
% To put the Starlink latency into context, 
For context, we also select speedtests originating from terrestrial serving-ISPs to capture mobile network traffic.
% We aim to compare the Starlink LEO network to consumer networks, ideally wireless mobile network (i.e., WiFi or cellular), by comparing it against the terrestrial serving-ISP infrastructure in that country.
%
We filter measurements from devices connected to the top-3 mobile network operators (MNOs) in each country~\cite{top3isps}.
%
% We identify \emph{mobile} devices by filtering for \texttt{android} and \texttt{iOS} tags within client ``environmentType'' in M-Lab measurements~\cite{mlab-measure}.
Note that our filteration results in a mix of wired and wireless access networks since M-Lab does not provide a way to distinguish between the two.
%
% We only consider measurements from the \textcolor{red}{past 12 months (i.e., April 2022--2023)} to ensure a fair comparison.
%
Our endpoint selection remains the same for both Starlink and terrestrial networks (see~\Cref{sec:methodology:global}).

\begin{comment}
First, we provide a global overview of the latencies achievable via Starlink in \Cref{fig:mlab_heatmaps}, showing the minRTT per-country.
%
Subsequently, we look at the global performance and explain our findings based on a selection of globally-distributed cities.
The minRTT distribution during tests from those cities over Starlink is shown in \Cref{fig:mlab_cdf_minrtt_glob}.
%
Finally, we characterize the latency by splitting the test results per-continent in \Cref{fig:mlab_cdf_regions} and investigate geographical differences and similarities%
\footnote{We picked the top-4/5 cities in terms of samples from each depicted region.
  In case of the US, we replaced Atlanta and Chicago with New York and Denver to also cover the countries central and eastern parts.
  Due to too few comparison points, Asia and Africa are not individually depicted.}%
.
%
We discuss our findings in the context of the available orbital shells and Starlink's terrestrial infrastructure in the areas depicted in \Cref{fig:mlab_shells}.
\end{comment}

\headline{Global View}
\Cref{fig:mlab_heatmaps} shows that, for a majority of countries, clients using terrestrial ISPs achieve better latencies over Starlink.
% \footnote{The first three bins of the depicted map were picked so the Starlink latencies are evenly distributed and the last three bins were picked to catch the outliers. The same bin definitions was then used with the top-3 serving ISPs of the respective countries.}%
% %
% We find that Starlink achieves more consistent median latencies globally ($\leq$ 100~ms), with outliers in Southeast Asia.
%
While the median latency of Starlink hovers around 40--50~ms, this distribution varies significantly across geographical regions.
% %
% The network performance of the LEO provider appears to be better in the United States and Europe.
%
% This is understandable since these regions emerged as ``early'' target markets of Starlink as the service was first offered in these countries~\cite{starlink-wiki} and they boast of significant availability of PoPs and gateways (see \cref{fig:mlab_gws}).
% This correlates with the significant availability of PoPs and gateways in these regions (see \Cref{fig:mlab_gws}) that were early target markets of Starlink~\cite{starlink-wiki}.
%
For instance, in Colombia, Starlink reports better latencies than established terrestrial networks.
Conversely, in Manila (The Philippines), Starlink's performance is notably inferior (\Cref{fig:mlab_cdf_minrtt_glob}).
%
% \Cref{fig:mlab_shells} shows the uneven distribution of GSes and PoPs around the world.
%The uneven distribution of GSs and PoPs (\Cref{fig:mlab_shells}) may be consequential as the USA, which achieves significantly lower latencies, boasts a robust ground infrastructure.
The uneven distribution of GSs and PoPs (\Cref{fig:mlab_shells}) may explain the latency differences;
the USA, which experiences significantly lower latencies, also boasts a robust ground infrastructure.
% For example, the USA has a dense ground infrastructure while the closest PoP from the Philippines is $\approx$ 3000~kms away.
%
% Similar trends are observable in Kenya and Mozambique, which are only served by the only available PoP in Nigeria.
We observe similar trends in Kenya and Mozambique where the closest PoP is located in Nigeria.
%
% We find that the performance of Starlink correlates with its ground infrastructure, i.e., GS and PoP, density (see \Cref{fig:mlab_probecities}) and discuss further results keeping this in perspective.
%
% Starlink is on par with the terrestrial ISPs in the US
%
% The result for US is surprising since several previous works have showcased significantly latencies over mobile networks~\cite{dang2021, 5g_look} ref.
%
%
% Our global median latency representation may be susceptible to skewness due to a few cities showcasing less-than-ideal performance.
% %
% To combat this, we plot the distributions of the minimum RTT (\texttt{minRTT}) reported during a speedtest over Starlink conducted within our measurement period from selected \emph{cities} globally in \Cref{fig:mlab_cdf_minrtt_glob}.
%
%

\begin{figure}[!t]
    \centering
	% \begin{subfigure}[htbp]{0.24\textwidth}
	% 	\includegraphics[width=\textwidth]{figures/mlab/state_2023-10-01/mlab_cdf_minrtt_5nacities_doctored.pdf}
	% 	\caption{\label{fig:mlab_cdf_minrtt_na} North America}
	% \end{subfigure}
	\begin{subfigure}[htbp]{0.23\textwidth}
		\includegraphics[width=\textwidth]{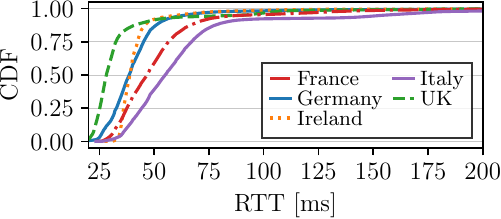}
		% \caption{Europe}
		\label{fig:mlab_cdf_minrtt_eu}
	\end{subfigure}
	\begin{subfigure}[htbp]{0.23\textwidth}
		\includegraphics[width=\textwidth]{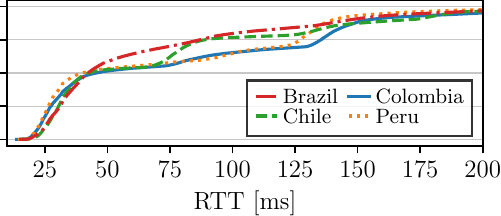}
		\label{fig:mlab_cdf_minrtt_sa}
		% \caption{South America}
	\end{subfigure}
	% \begin{subfigure}[htbp]{0.24\textwidth}
	% 	\includegraphics[width=\textwidth]{figures/mlab/state_2023-10-01/mlab_cdf_minrtt_5aucities_stripped.pdf}
	% 	\caption{\label{fig:mlab_cdf_minrtt_au} Oceania}
	% \end{subfigure}
	\vspace*{-2em}
	\caption{\label{fig:mlab_cdf_regions} 
	% Distributions of minimum reported RTTs during M-Lab measurements from selected cities in North America, Europe, South America and Oceania, respectively.
	Distributions of M-Lab minRTTs over Starlink from select cities in Europe and South America, respectively.
	}
        % \vspace*{-2em}
        \vspace*{-5mm}
\end{figure}

\headline{Well-Provisioned Regions}
% \headline{Latencies in well-provisioned regions}
% \headline{Latency Performance in the US}
Even though a significant portion of global Starlink measurement samples originate from Seattle ($\approx 10\%$), 
% which is also the location of Starlink's headquarters.
%
% Despite a large number of samples, measurements from Seattle 
the region shows consistently low latencies, with the 75\textsuperscript{th} percentile well below 50~ms (\Cref{fig:mlab_cdf_minrtt_glob}).
%
% This can be attributed to several reasons, e.g., 
Contributing factors can be dense GS availability or service prioritization for Starlink's headquarters.	
% home area for research.
%
We also observe that Starlink performance is fairly consistent across the USA, confirming that Seattle is not an anomaly but the norm (see \Cref{fig:mlab_cdf_minrtt_na} in \Cref{app:global}).
% verify the consistency of Starlink across the USA by comparing the Starlink latencies reported from other US cities in \Cref{fig:mlab_cdf_minrtt_na}. %(selection based on the number of measurement samples).
% %
% The representation confirms that Seattle is not an outlier but that all US cities closely follow the same latency distribution: The 90\textsuperscript{th} percentile of almost all measurements stays below 50ms.
%
This highlights the LEO network's potential to bridge Internet access disparities, 
% achieve Internet access equality independent of factors such as geographical location and income, 
which currently affects the quality of terrestrial Internet in the USA~\cite{inequity-california,inequity-chicago}.
%
% This also explains our initial observation of Starlink outperforming terrestrial ISPs in the USA.
%
Europe is also relatively well covered with GSs but hosts only three PoPs that are in the UK, Germany, and Spain.
Proximity to the nearest PoP correlates strongly with minRTT performance in \Cref{fig:mlab_cdf_regions} -- 
Dublin, London, and Berlin exhibit comparable latencies to the US, while for Rome and Paris, the 75\textsuperscript{th} percentile is $\approx$~20~ms longer.
Unlike US, Starlink latencies in EU has longer tail, often surpassing 100~ms.
%The performance discrepancy raises the question if also other factors influence the performance in Rome and Paris but it is reasonable to conclude that geographically close PoPs are important for low latency.

\headline{Under-Provisioned Regions}
% \Cref{fig:mlab_heatmaps} suggests that Starlink has the potential to benefit regions with inconsistent terrestrial mobile network deployments~\cite{speedtest-index}.
% %
% Tests from Colombia exhibit $\approx$~10--20\% reduced latency over Starlink compared to terrestrial, 
% % serving-ISPs,
% with median minRTTs almost equal in Mexico (Starlink being 1~ms lower).
% Starlink: 83~ms, terrestrial serving-ISPs: 84~ms).
Starlink's superior performance in Colombia hints at its potential for connecting under-provisioned regions.
However, \Cref{fig:mlab_cdf_regions} shows that Starlink in South America (SA) trails significantly behind the US and Europe, with the 75\textsuperscript{th} percentile exceeding 100~ms and tail reaching 200~ms.
%
% We observe a similar trend for Australia as well as almost all cities: They show a minRTT at the 75th-percentile within 50~ms while the rest exhibit significantly higher latencies exceeding even 200~ms (see \Cref{fig:mlab_cdf_minrtt_au}).
%
% Similarly, the RTT performance in Oceania ranks behind the US and Europe -- minRTT in \Cref{fig:mlab_cdf_minrtt_au} at the 75\textsuperscript{th} percentile is significantly larger than 50~ms, and the tail latencies exceed 200~ms.
Oceania also shows similar performance (see \Cref{fig:mlab_cdf_minrtt_au} in \Cref{app:global}).
%
% \textcolor{red}{Since these regions are covered by the same dense 53$\degree$ inclination orbits, the potential causes for degraded performance can be: (i) lower density of GSes and PoPs in these regions; or (ii) congestion in the terrestrial network between the PoPs and endpoint servers.
% %
% New Zealand has multiple GSes and a PoP close to its most populous city Auckland; its performance is noticeably better than in Australian cities like Perth and Sydney.
% %
% We cannot extract the access (satellite bent-pipe) latency from the end-to-end M-Lab measurements and, therefore, investigate this further from RIPE Atlas probes in \Cref{subsec:ripeatlas}.} % \Cref{sec:methodology:atlas}
% We attempt to understand the 
By extracting the share of 
% isolate the latency contribution of the 
satellite vs.\ terrestrial path (i.e PoP $\leftrightarrow$ M-Lab servers, see \Cref{fig:mlab_heatmap_starlink_latency_fraction} in \Cref{app:global})\footnotemark, we find that the majority of SA Starlink latency is due to the bent-pipe. 
% \fotenotetext{We subtract latencies reported by \texttt{reverse traceroute}s from end-to-end RTT.}
\footnotetext{We subtract the latency to the Starlink PoP reported by M-Lab's \texttt{reverse traceroute}s from the end-to-end TCP minRTT.}
% However, in regions with mature mobile network infrastructure deployment, Starlink does not provide any significant benefit (see Brazil, Chile, etc.).
%
% An outlier from \cref{fig:mlab_heatmaps} is the United States, where Starlink reportedly achieves slightly better latencies over terrestrial serving ISPs.
% %
% To gain further insight, we plot the distribution of Starlink latencies from select cities across the globe in \cref{fig:mlab_cdf_global}.
%
In contrast, latencies from Mexico and Africa (except Nigeria) show terrestrial influence, which we allude to non-optimal PoP assignments by Starlink routing.

We also observe an interesting impact of ground infrastructure in the Philippines, where a local PoP was deployed in May 2023.
Prior to this, Starlink traffic from the country was directed to the nearest Japanese PoP, traversing long submarine links to circle back to the geographically closest M-Lab server in Philippines -- evident from additional 50--70~ms RTT in \Cref{fig:mlab_cdf_minrtt_phjp} in \Cref{app:global} for Philippines users to reach in-country vs. Japanese M-Lab servers.
However, post-May 2023, the latencies to in-country servers reduced by 90\% as the traffic was routed via the local PoP.
%
% We improve this rudimentary analysis in \cref{sec:last-mile} through our targeted \texttt{traceroute} from Starlink terminals, which allows us accurate segment latencies over the end-to-end path.
%
Despite such artifacts, Starlink shows an evident trend towards more consistent sub-50~ms latencies globally over the past 17 months,
% latency evolution over the past 17 months shows a trend towards more consistent sub-50~ms, 
specifically evident in Sydney
(\Cref{fig:mlab_temporal_2022-2023_minrtt} in \Cref{app:global}).
%
% In addition to launching more satellites in orbit that effectively handle growing subscriber load~\cite{starlink-wiki}, increasing ground station deployment and improved routing~\cite{kassem2022} may
% Several factors may have caused this, such as (i) additional GS and PoP deployments, (ii) more operational satellites (see \Cref{tbl:starlink_orbits} in \Cref{app:orbits}), and (iii) improved routing~\cite{kassem2022}.
%
We conclude that while Starlink slightly lags behind terrestrial networks today, the gap 
% performance gap between Starlink and \emph{today's} wireless networks 
will continue to shrink as the ground (and satellite) infrastructure expands.
% expands; a detailed comparison to the \textit{evolving} 5G and beyond cellular networks is a topic for future study.
% We conclude that the performance gap between Starlink terrestrial wireless networks will continue to shrink with a continued expansion of the supporting ground infrastructure.
% \textcolor{red}{JO: Yes, but any Starlink Internet access link satellite will need to go via 2x2x500+ = 2000km distance, so some 10ms speed of light delay for shells further up.  With all the 5G/6G talk on 1ms access latency to the next cell tower, this may be tough competition as also the terrestrial networks in well-provisioned areas evolve. So, I would try to word this more carefully as in (maybe too long): We conclude that the performance gap between Starlink and \textit{today's} wireless networks will continue to shrink as Starlink's ground (and satellite) infrastructure expand; a comparison to the \textit{evolving} 5G and beyond cellular networks obviously remains for future study.}

\begin{figure}[t!]
    \centering
	\includegraphics[width=\columnwidth]{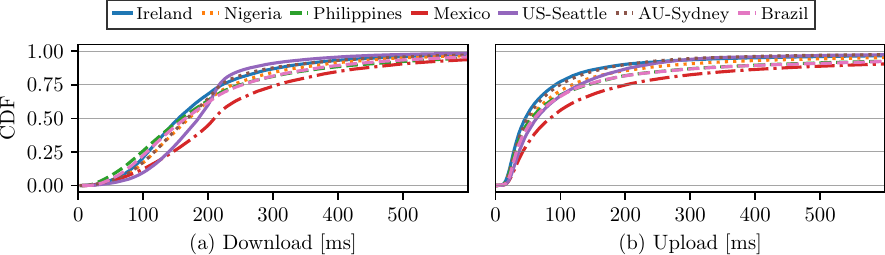}%
    \vspace*{-1em}
    \caption{\label{fig:mlab_bufferbloat}\label{fig:mlab_bufferbloat:download}\label{fig:mlab_bufferbloat:upload}
    RTT inflation (\texttt{maxRTT}-\texttt{minRTT}) during M-Lab Starlink speedtests per continent: (a) download, (b) upload traffic.
    \vspace*{-2.5em} % pull up latency-under-load headline}
    }
\end{figure}

\begin{figure}[t!]
  \centering
  \includegraphics[width=\columnwidth]{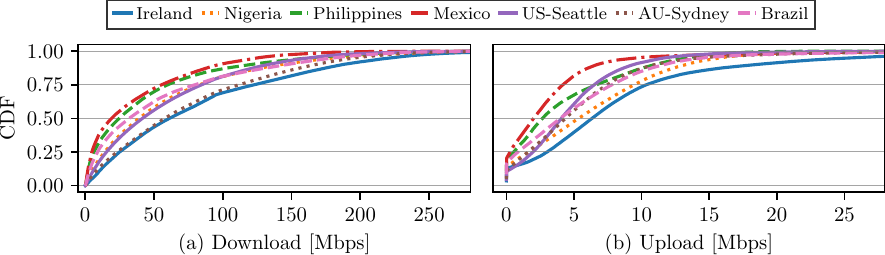}%
  \vspace*{-1em}
  \caption{\label{fig:mlab_cdf_goodput_glob} Distribution of median (a) download and (b) upload goodput over Starlink from selected cities globally.}
  \vspace*{-1.8em}
\end{figure}

\headline{Latency Under Load}
Recent findings suggest that Starlink may be susceptible to \emph{bufferbloat}~\cite{bufferbloating, bufferbloat-icc2017}, wherein latencies during traffic load can increase significantly due to excessive queue buildups~\cite{michel2022}.
%
% To verify this from a global perspective
To explore this globally, we evaluate the RTT inflation, i.e., the difference between the maximum and minimum RTT observed during a speed test.
% data transmissions over the Starlink network.
%
% We estimate the ``inflated RTT'' as the difference between the maximum and lowest RTT observed within a single test.
%
% Recall that M-Lab \texttt{ndt7} measurements use a single 10~s TCP BBR connection for both upload and download.
% with BBR during download tests and a client-chosen congestion control algorithm during uploads.
% While BBR aims to keep in-network queues low, recent studies show that this goal is not always achieved~\cite{bbr-bufferbloat}.
% Therefore, we believe that RTT inflation is a reasonable approximation of assessing excessive queueing delays.
%
% \Cref{fig:mlab_bufferbloat} confirms presence of bufferbloating within Starlink globally.
\Cref{fig:mlab_bufferbloat} reveals significant delay inflation under load 
% within Starlink globally.
% as during active downloads
%
as during active downloads, Starlink experiences $\approx$~2--4$\times$ increased RTTs, reaching almost 400--500~ms (\Cref{fig:mlab_bufferbloat:download}a).
While such inflations are consistent across \emph{all} Starlink service areas, they are more prominent in regions with subpar baseline performance, e.g., Mexico.
Note that the Starlink latency under load is not symmetric.
The 60\textsuperscript{th} percentile of RTT during uploads increases to $\leq$~100~ms globally (see \Cref{fig:mlab_bufferbloat:upload}(b)) compared to $\approx$~200~ms during downloads.
We observe similar behavior while conducting \texttt{iperf} over our controlled terminals.
Possible explanations can be queue size differences at the Dishy (affecting uploads), the ground station (affecting downloads), or satellites (impacting both).
%
% Note that our findings are in stark contrast to a previous study, which found more pronounced latency inflation during uploads instead of downloads~\cite{michel2022}.
% %
% This difference could be due to (potential) sampling bias since the authors infer their takeaways from one week of measurements over a single Starlink terminal located in Europe
% %
% while we leverage $>$ 8M crowdsourced measurements globally.
% \textcolor{red}{JO: Just checking: could we have measurement platform-induced biases to mention here?}
%
% {\color{red}The increased loss rates during uploads compared to downloads (see fig XXX) further strengthen our suspicion since smaller-sized queues are more likely to drop packets.
% %
It is also plausible that Starlink employs active queue management (AQM) techniques~\cite{adams2012active} to moderate uplink latencies under congestion.
This approach, however, may adversely impact applications that demand both high bandwidth and low latency -- which we explore in \cref{sec:cloud-gaming}.
\headline{Goodput}
% Here we will talk about the throughput achievable. KDE plots, etc.
\Cref{fig:mlab_cdf_goodput_glob} shows Starlink download and upload goodputs from speed tests globally.
% We analyze the TCP goodput, that excludes retransmission, as we see it to be more important to the performance observed by an end-user of Starlink.
%
% Compared to the significant differences in latencies between the selected cities (see \Cref{fig:mlab_cdf_minrtt_glob}), the goodput distributions are more homogeneous.
Unlike latencies (\Cref{fig:mlab_cdf_minrtt_glob}), the goodput distributions appear relatively homogeneous. 
Most Starlink clients achieve $\approx$~50--100~Mbps download and $\approx$~4--12~Mbps upload rates at the 75\textsuperscript{th} percentile.
We also do not find any correlation between baseline latencies (see \Cref{fig:mlab_cdf_minrtt_glob}) and upload/download goodput, evident from the contrasting cases of Dublin and Manila.
% , where the latter achieves comparable  goodputs despite significantly higher latency.
% that show inversely proportional trend between the two  .
%
% On the one hand, a small/large minRTT yields large/small download and upload goodputs in the case of Dublin/Manila.
% This is expected, since TCP increases its sending rate more quickly on lower-RTT connections, can work with a smaller receive window, and can react more quickly to packet losses.
%
However, we observe an inverse correlation between loss rates and goodputs; increasing from 4--8\% at the 75\textsuperscript{th}-percentile (see \Cref{fig:mlab_cdf_lossrate-down_global} in \Cref{app:global}).
%
% On the other hand, 
Seattle, notable for its latency performance,
% which performed remarkably well from latency perspective, 
records average goodputs. 
% falls in the middle.
% On the other hand, Sydney has a large minRTT for the top-25\% of tests but a comparatively high download goodput.
%
% However, cities that report consistently low latencies (i.e, Dublin, Seattle, and Lagos) do not nessarily also achieve high throughputs.
% %
% While Seattle appears to be an outlier with lower goodputs despite outperforming other selected cities from a latency perspective.
% %
% Being the most populous from measurement density viewpoint, the trend may be due to 
% % a large number of users in the area which could necessitate 
% Starlink internal throttling/load balancing policies to avoid congestion on the shared bent-pipe~\cite{starlink-fairuse}.
Considering high measurement density from this region, the trend might be due to Starlink's internal throttling or load-balancing to prevent congestion~\cite{starlink-fairuse}.
%
% We attempt to understand this better by comparing the evolution of goodputs achieved from the selected cities over the last 17 months (see \cref{fig:mlab_temporal_global} in \Cref{app:global}).
%
We also find that over the past 17 months, Starlink goodputs have stabilized rather than increased, with almost all geographical regions demonstrating similar performance (shown in \Cref{fig:mlab_temporal_global} in \Cref{app:global}). 

\begin{tcolorbox}[title=\textit{Takeaway \#1}, enhanced, breakable]
    % Starlink is on the path to becoming a global ISP with increasing service coverage and dense satellite availability. However, its current offering is not yet competitive, especially in Mexico and the Philippines, due to dense GS and PoP on-ground deployment requirements for bent-pipe operation. Regardless, Starlink clients globally observe a long tail in latency and significantly increased RTTs during traffic load, likely due to bufferbloating. However, we notice coherence in latency and goodput performance across different regions served by the same orbit over time, which is a promising sign for Starlink's future.
  Starlink exhibits competitive performance to terrestrial ISPs on a global scale, especially
  % Starlink network performance is comparable to terrestrial ISPs 
  in regions with dense GS and PoP deployment.
  However, noticeable degradation is observable in regions with limited ground infrastructure.
  % in achieves significantly degraded higher baseline latencies and latency under load 
  % where infrastructure is limited.
  Our results further confirm that Starlink is affected by bufferbloat.
  Starlink appears to be optimizing for consistent global performance, albeit with a slight reduction in goodput.
%    likely due to increasing subscriber base.
  %
  % ISLs may decouple Starlink's dependence on ground infrastructure; however, its performance impact requires further investigation.
  % Manila is an extreme example where distant PoP locations with regards to the destination server cause significantly worse round-trip times and also goodput.
  % On the other hand, the evolution of Starlink is resulting in lower and less variable network delays, likely fueled by more ground infrastructure and operational satellites.
  % %
  % Goodput globally is heading downwards though, presumably due to increase in the number of users.
\end{tcolorbox}

% \begin{figure}[htbp]
% 	\begin{subfigure}[htbp]{0.4\textwidth}
% 		\includegraphics[width=\textwidth]{figures/mlab/temp_2022-02_boxplot_over_time.png}
% 		\caption{\label{fig:temp_geomagstorm} \textcolor{red}{Temporary data dump. [NM]: All cities wont work. TODO: for each city a separate CDF} Boxplot over time showcasing time ranges where a geomagnetic storm occurred in february 2022}
% 	\end{subfigure}
% 	\caption{temporary dump}
% \end{figure}

% \begin{enumerate}
% 	\item [x] World map of MLab measurement IPs geolocated
% 	      \subitem maybe with longitudinal/latitudinal histogram
% 	\item [x] MLab probe density with launch background
% 	      \subitem maybe with longitudinal/latitudinal histogram
% 	\item [x] Global MinRTT and throughput CDF (TODO: fix sizes)
% 	\item [x] EU MinRTT CDF
% 	\item [x] SA MinRTT CDF
% 	\item [x] US MinRTT CDF
% 	\item [x] AU MinRTT CDF
% 	\item [x] Boxplot over Time for selected cities
% 	\item [x] Comparison of MNO and Starlink (what plot exactly tbd)
% 	\item [ ] weather anomalies impact - \url{https://en.wikipedia.org/wiki/December_2022_North_American_winter_storm}
% \end{enumerate}

\section{Real-time Application Performance} \label{sec:cloud-gaming}

\begin{figure}[t]
  \centering
  \includegraphics[width=\columnwidth,page=2,trim={0 3mm 0 0}]{figures/zoom_xtime.pdf} % legend
  % \vspace{-2mm}
  \includegraphics[width=\columnwidth,page=1]{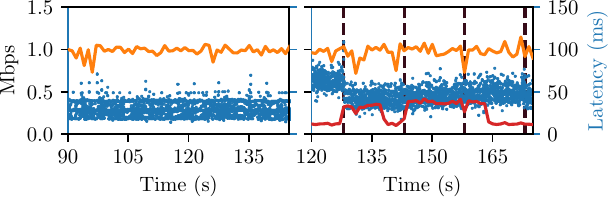} % figure
  \vspace*{-2em}
  \caption{\label{fig:zoom-xtime} Uplink Zoom traffic over a terrestrial (left) and Starlink (right). Vertical lines show 15~s reconfigurations.}
  \vspace*{-2em}
\end{figure}

While the global Starlink performance in \Cref{sec:global} is promising for supporting web-based applications, it does not accurately capture the potential impact of minute network changes caused by routing, satellite switches, bufferbloating, etc., on application performance.
Real-time web applications are known to be sensitive to such fluctuations~\cite{imc2021zoom, michel2022, iqbal2021}.
% therefore necessitating further investigation.
%
In this section, we examine the performance of Zoom and Amazon Luna cloud gaming over Starlink %deployed in \textcolor{red}{two} controlled geographical locations 
(see \Cref{sec:methodology:gaming} for measurements details).
%
% We holistically capture the requirements of the majority of real-time Internet-based applications, as cloud gaming imposes a strict latency control loop along with high downlink bandwidth, whereas Zoom utilizes uplink and downlink capacity simultaneously.
This allows us to assess the suitability of the LEO network to meet the requirements of the majority of real-time Internet-based applications, as both applications impose a strict latency control loop.
Cloud gaming necessitates high downlink bandwidth, while Zoom utilizes uplink and downlink capacity.
% \begin{figure}[t]
%   \centering
%   \includegraphics[width=.6\columnwidth,page=2,trim={0 2mm 0 0}]{figures/zoom_boxplots.pdf} % legend
%   \includegraphics[width=\columnwidth,page=1]{figures/zoom_boxplots.pdf} % figure
%   \vspace{-7mm}
%   \caption{\label{fig:zoom-boxplots} Zoom performance over Starlink and a terrestrial link. \textcolor{red}{Preliminary data.}}
% \end{figure}

% \begin{figure}[t]
%   \centering
%   \includegraphics[width=.8\columnwidth,page=2,trim={0 2mm 0 0}]{figures/gaming_xtime.pdf} % legend
%   % \vspace{-2mm}
%   \includegraphics[width=\columnwidth,page=1]{figures/gaming_xtime.pdf} % figure
%   \vspace{-7mm}
%   \caption{\label{fig:gaming-xtime} Cloud gaming over 5G (left) and Starlink (right).}
%   % \vspace{-0.25cm}
% \end{figure}

% \begin{figure}[t]
%   \centering
%   \includegraphics[width=0.75\columnwidth,page=2]{figures/gaming_xtime.pdf}%
%   \vspace{-2mm}
%   \includegraphics[width=\columnwidth,page=1]{figures/gaming_xtime.pdf}
%   \vspace{-6mm}
%   \caption{\label{fig:gaming-xtime} A sample of cloud gaming over the terrestrial connection (left), cellular 5G (middle), and Starlink (right).}
%   \vspace{-0.25cm}
% \end{figure}

\headline{Zoom Video Conferencing}
\Cref{fig:zoom-xtime} shows samples from Zoom calls conducted over a high-speed terrestrial network and over Starlink.
The total uplink throughput over Starlink is slightly higher,
which we trace to FEC (Forward Error Correction) packets that are frequently sent in addition to raw video data (on average 146\stddev{99}~Kbps vs. 2\stddev{2}~Kbps over terrestrial).
% FEC data is almost never sent over the terrestrial connection.
The frame rate, inferred from the packets received by the Zoom peer, does not meaningfully differ between the two networks ($\approx$~27~FPS).
Note that, since Zoom does not saturate the available uplink and downlink capacity, it should not be impacted by bufferbloating.
Yet, we observe a slightly higher loss rate over LEO, which the application combats by proactively utilizing FEC.
% The proactive error correction scheme seems to mitigate Starlink's supposedly higher packet loss rate.
%
The uplink one-way delay (OWD) over Starlink is higher and more variable compared to the terrestrial connection (on average 52\stddev{14}~ms vs. 27\stddev{7}~ms).
All observations also apply to the downlink except that Starlink's downlink latency (35\stddev{11}~ms) is similar to the terrestrial connection (32\stddev{7}~ms).
Our analysis broadly agrees with~\cite{zhao2023real} but
our packet-level insight reveals bitrate fluctuations partly caused by FEC.
Further, our Starlink connection was more reliable and we did not experience second-long outages.

Interestingly, we observe that the Starlink OWD often noticeably shifts at interval points that occur at 15~s increments.
Further investigation reveals the cause to be the Starlink \emph{reconfiguration interval}, which, as reported in FCC filings~\cite{starlink2021petition}, is the time-step at which the satellite paths are reallocated to the users.
Other recent work also reports periodic link degradations at 15~s boundaries in their experiments, with RTT spikes and packet losses of several orders~\cite{pan2023measuring, kassem2022, tanveer2023making}.
We explore the impact of reconfiguration intervals and other Starlink-internal actions on network performance in \Cref{sec:last-mile}.

% that is confirmed by the median calculated over 6 hours of calls over each network:
%

\begin{table}[!t]
\newcommand{\hlcell}{\cellcolor{gray!25}}
\tabcolsep=3pt % Reduce the inter-column space
\begin{tabular}{@{}lrrr@{}}
\toprule
                     & Terrestrial          & Cellular             & Starlink             \\ \midrule
Idle RTT (ms)        & 9                    & \hlcell46                   & 40                   \\
Throughput (Mbps)    & 1000                 & \hlcell150                  & 220                  \\
\midrule
% Frames-per-second    & 58.9\stddev{1.51}    & 59.26\stddev{1.68}   & 58.95\stddev{1.63}   \\ % mean
Frames-per-second    & 59\stddev{1.51}    & 59\stddev{1.68}   & 59\stddev{1.63}   \\ % .ren_fps, metrics_{}
%
% Bitrate (Mbps)       & 23.02\stddev{0.38}   & 20.92\stddev{4.24}   & 22.05\stddev{2.16}   \\ % mean
Bitrate (Mbps)       & 23.08\stddev{0.38}   & 22.82\stddev{4.24}   & 22.81\stddev{2.16}   \\ % .rx_bps, metrics_{}
Time at 1080p (\%)   & 100                  & \hlcell94.11                & 99.45                \\ % .height_1080_rel, metrics_{}
%
% Drops (frames/min)   & 0                    & 0.96\stddev{5.08}    & 0.41\stddev{1.95}    \\
%
% Freezes (ms/min)     & 0                    & 47.5\stddev{220.34}  & 40.06\stddev{119.74} \\ % mean
Freezes (ms/min)     & 0\stddev{0}          & \hlcell0\stddev{220.34}  & 0\stddev{119.74} \\ % .total_freezes_duration_ms, metrics_resampled_{}
%
% Inter-frame (ms)     & 16.78\stddev{3.65}   & 16.72\stddev{11.1}   & 16.85\stddev{6.76}   \\ % mean
Inter-frame (ms)     & 17\stddev{3.65}   & \hlcell18\stddev{11.1}   & 16\stddev{6.76}   \\ % .interframe_delay_ms, metrics_{}
\midrule
% Game delay (ms)      & 138.42\stddev{19.79} & 166.94\stddev{23.55} & 171.42\stddev{23.12} \\ % mean
Game delay (ms)      & 133.53\stddev{19.79} & 165.82\stddev{23.55} & \hlcell167.13\stddev{23.12} \\ % .gaming_delay, metrics_{}
%
% \hspace*{1em}Server (ms)          & 89.1\stddev{2.62}    & 100.87\stddev{4.66}  & 94.07\stddev{3.03}   \\
%
% \hspace*{1em}RTT (ms)             & 22.98\stddev{13.41}  & 43.45\stddev{17.06}  & 51.2\stddev{16.28}   \\ % mean
\hspace*{1em}RTT (ms)             & 11\stddev{13.41}  & 39\stddev{17.06}  & \hlcell50\stddev{16.28}   \\ % .rtt, metrics_{}
%
% \hspace*{1em}Jitter buffer (ms)   & 15.89\stddev{3.27}   & 12.2\stddev{1.33}    & 15.86\stddev{3.35}   \\ % mean
\hspace*{1em}Jitter buffer (ms)   & 15\stddev{3.27}   & 12\stddev{1.33}    & \hlcell15\stddev{3.35}   \\ % .jb_delay, metrics_{}
%
% \hspace*{1em}Decoding (ms)        & 0.45\stddev{0.5}     & 0.43\stddev{0.5}     & 0.29\stddev{0.46}    \\
% \hspace*{1em}Rendering (ms)       & 10\stddev{0}         & 10\stddev{0}         & 10\stddev{0}         \\
\bottomrule
\end{tabular}
\caption{\label{tab:gaming-metrics}
    Luna gaming results over 150~mins playtime.
  Values denote median\stddev{SD} and the worst performer is highlighted.
  % Frame drops and freezes are calculated over 60-second windows.
  % The link metrics are reference points obtained using Ookla Speedtest.
  \vspace*{-2.5em}
}
\end{table}

% \begin{figure}[t]
%   \centering
%   \includegraphics[width=\columnwidth]{figures/gaming_delay_parts.pdf}
%   \caption{\label{fig:gaming-delay-parts} The overall game delay split into its components.
%     Only the network interface was varied between the tests, which affects the network and jitter buffer delays.
%     \textcolor{red}{HC: replace ``terrestrial'' with ``wired''? The ``cellular'' link is also in a sense ``terrestrial'', right?}
%     \textcolor{red}{HC: Remove this figure (1) since the interesting values (can be presented in the table, and (2) since I can’t explain why the ``server'' part varies between the tests?}}

%   \end{figure}

\begin{figure}[t]
  \centering
  \includegraphics[width=.8\columnwidth,trim={0 3mm 0 0},page=2]{figures/gaming_xtime.pdf} % legend
  % \vspace{-2mm}
  \includegraphics[width=\columnwidth,page=1]{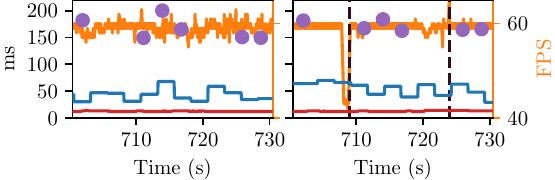} % figure
  \vspace{-2.5em}
  \caption{\label{fig:gaming-xtime} Cloud gaming over 5G (left) and Starlink (right). Vertical dashed lines show Starlink reconfiguration intervals.}
  \vspace*{-2em}
\end{figure}

\headline{Amazon Luna Cloud Gaming}
\Cref{tab:gaming-metrics} shows 150 minutes of cloud gaming performance over terrestrial, 5G cellular, and Starlink networks.
Overall, all networks realized close to 60 FPS playback rate at consistently high bitrate ($\approx$ 20~Mbps).
Starlink lies in between the better-performing terrestrial and cellular in terms of 
% While Starlink exhibited more variability than the wired link, it delivered the video more stably than the cellular link in terms of 
bitrate fluctuations, frame drops and freezes%
\footnote{Freeze is when the inter-frame delay (IFD) is larger than $\text{max}(3\times$$\text{IFD}, \text{IFD} + 150)$.}.
Starlink exhibits the highest game delay, i.e., the delay experienced by the player between issuing a command and witnessing its effect.
% shows a more noticeable distinction between the wired and the wireless access technologies.
% \autoref{fig:gaming-delay-parts} decomposes the game delay into its parts and highlights the impact of the network latency.
Specifically, the wired network delivers the visual response about 2 frames ($\approx$~33~ms) earlier than both 5G and Starlink.
%
% The majority of that difference comes from a lower RTT but we also notice considerable differences in the server-processing time.
% %
% That part is calculated as the difference between the overall game delay and the other four depicted game delay contributors.
% The game engine and video encoding performance on the cloud server should be independent from the tested network connection, similar to the decoding and rendering stages.
% However, we observe that it contributes more to the wireless game delay compared to the wired cases.
% We assume that packet losses and reordering on the uplink, where player commands are sent, cause delays in generating frames that contain the visual response to the player's action.
%
While examining the gaming performance over time, we observe occasional drops to <~20~FPS over Starlink (see \Cref{fig:gaming-xtime}), that coincide with Starlink's reconfiguration interval.
These fluctuations are only visible at sub-second granularity and, hence, are not reflected in global performance analysis (\Cref{sec:global}).

Despite these variations, Starlink's performance remains competitive with 5G, highlighting its potential to deliver real-time application support, especially in regions with less mature cellular infrastructure.
Note, however, that our Starlink terminal was set up without obstructions and the weather conditions during measurements were favorable to its operation~\cite{ma2022network}.
% while we carried out the stationary tests.
Different conditions, especially mobility, may change the relative performance of Starlink and cellular, which we plan to explore further in the near future.
%
%As such, our experiments highlight the potential of Starlink's capabilities to deliver real-time support in regions with less mature cellular infrastructure remains to be investigated.
% As such, our experiments also present a biased view, and we plan to evaluate Starlink as a key enabler for latency-critical applications, such as cloud gaming, in regions with sub-optimal terrestrial Internet connectivity in the near future.
% On the other hand, being the only consumer-targeted LEO satellite network, Starlink can potentially become a monopoly within this sector and degrade its performance over time as its global subscribers increase.
% As the demands for real-time applications increase, we expect the impact of the reconfiguration interval to become significantly more apparent in the end-user performance.
% For example, autonomous drones are an application that seem, at first glance, to be well-suited to Starlink, particularly if used in remote areas.
% However, the combination of the strict low-latency requirements along with high throughput demands (especially for multi-camera drones) has the potential to saturate the connection to the point where the reconfiguration interval may become problematic.

\vspace*{-0.2em}
\begin{tcolorbox}[title=\textit{Takeaway \#2}, enhanced, breakable]
    % Submitted:
    % Starlink is competitive with the current 5G deployment for supporting demanding real-time applications.
    % Camera-ready:
    Starlink's performance is competitive with the current 5G deployment for supporting demanding real-time applications.
    %
    % compared to terrestrial networks, but its continued ability to do so in the future remains to be seen.
    % and is competitive with current 5G deployments.
    We also observe that Starlink experiences regular performance fluctuations every 15s linked to its reconfiguration intervals.
    % While such internal black-box parameters influence performance to some degree, application-specific corrective measures, such as FEC, help overcome these artifacts.
    While these internal black-box parameters do influence performance to a certain extent, application-specific corrective measures, like FEC, are effective in mitigating these artifacts.
    % Proactive error correction schemes can mitigate Starlink's performance shortcomings.
    
  \end{tcolorbox}

\section{Dissecting the Bent-Pipe} \label{sec:last-mile}

% While our analysis in the previous section provided insights on the end-to-end performance of Starlink globally,
% it did not investigate the breakup and factors that can further affect its performance.
% it did not investigate the individual parts that affect its behavior.
%
% Recent studies have hinted at the bent-pipe (see \Cref{fig:starlink-bentpipe}) as the potential bottleneck within the Starlink network~\cite{michel2022, kassem2022}.
%
% Researchers have also conjectured switching to inter-satellite links (ISLs) instead~\cite{leo-delay, heaven, gearing}, which promises lower latencies and expanded connectivity areas but at the cost of increased variability~\cite{leo-route, l2-d2}.
% Researchers have also explored the effect of enabled inter-satellite links (ISLs)~\cite{leo-delay, heaven, gearing}, which
% promise lower latencies and expanded connectivity areas but at the cost of increased variability~\cite{leo-route, l2-d2}.
% %
% While Starlink has alluded to actively supporting ISLs before the end of 2023~\cite{starlink-isl}, the significant GS and PoP deployments hint at a ``hybrid'' operation where the traffic traverses some ISL hops to a better positioned GS instead of the closest one~\cite{hypatia}.

% In this section, we 
% dig deeper into Starlink's traffic engineering and 
We now attempt to uncover Starlink's behind-the-scenes operations and their impact on network performance.
% In this section, we examine the \emph{current state} of Starlink's bent-pipe behavior.
%
We follow a two-pronged approach to undertake this challenge.
Our longitudinal \texttt{traceroute} measurements over RIPE Atlas accurately isolate the bent-pipe (terminal-to-PoP) global performance, allowing us to correlate it with parameters like ground station deployment, satellite availability, etc. (\cref{subsec:ripeatlas}). 
% \Cref{app:starlink-path} discusses which hops in these measurements we consider for the bent-pipe latencies.
%
We then perform controlled, high-resolution experiments over Starlink terminals deployed in two EU countries to zoom in on bent-pipe operation and highlight traffic engineering signatures that may impact application performance (\Cref{subsec:controlled}).
%% First add the global perspective

\vspace*{-1em}
\subsection{Global Bent-Pipe Performance}\label{subsec:ripeatlas}
\begin{figure}[t!]
    \centering
    \includegraphics[width=0.9\columnwidth]{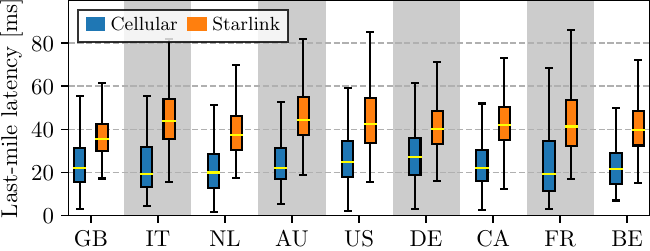}
    \vspace*{-1em}
    \caption{\label{fig:ripe_countries}
      Last-mile latencies for different countries.
      ``Starlink'' denotes satellite bent-pipe over RIPE Atlas while
      ``Cellular'' wireless access from Speedchecker~\cite{dang2021}.}
      \vspace*{-1.5em}
\end{figure}

% \vspace{-0.25cm}
\noindent\textbf{\emph{Starlink vs. Cellular Last-mile}}
We contrast our end-to-end M-Lab and real-time application analysis by comparing the Starlink bent-pipe  latencies from RIPE Atlas \texttt{traceroute}s to cellular wireless last-mile (device-to-ISP network) access.
Given the under-representation of cellular probes in RIPE Atlas, we augment our dataset with recent measurements from~\citet{dang2021}, which leveraged 115,000 cellular devices worldwide over the Speedchecker platform.
%  to analyze the performance of cellular networks worldwide.
%
%
\Cref{fig:ripe_countries} presents a comparative analysis of Starlink and cellular last-mile across countries common in both datasets. 
% illustrates the latency of the last-mile recorded from different countries with available Starlink Atlas probes.
% It also compares these results with last-mile latency measurements from cellular-connected Speedchecker probes that are derived from~\cite{dang2021}, where the last-mile denotes the latency between the user device and the cellular tower. 
%
% values of the probes connected to the Starlink network in the different countries. 
Consistent with our previous findings, we find that the Starlink bent-pipe latencies fall within 36--48~ms, with the median hovering around 40~ms for almost all countries.
%  This is illustrated in ~\Cref{fig:ripe_tracert_last_mile_global_abs} in \Cref{sec:global:latency}.
%
% values for the last mile access latency lie within a range of 10-80ms, with the majority of the values (25-75th percentile) for all countries ranging from 30-55 ms. 
%
Similarly, we find consistent cellular last-mile latencies across all countries, but almost 1.5$\times$ less than Starlink.
Recent investigations~\cite{pruning-hotnets2020} report similar access latencies over WiFi and cellular networks.
The bent-pipe latencies also corroborate our estimations in \Cref{sec:global} that the terminal $\leftrightarrow$ PoP path is the dominant contributor to the end-to-end latency.
% provided sufficient ground infrastructure is available.
%
Out of the 21 countries with Starlink-enabled RIPE Atlas probes, the only exceptions where the bent-pipe latency is significantly higher ($\approx$ 100~ms) are the Virgin Islands (US), Reunion Islands (FR), and Falkland Islands (UK).
Correlating with \Cref{fig:mlab_probecities}, we find that Starlink neither has a GS nor a PoP in these regions, which may result in traffic routing over ISLs to far-off GS leading to longer bent-pipe latencies.

\begin{figure}[t!]
	\includegraphics[width=0.9\columnwidth]{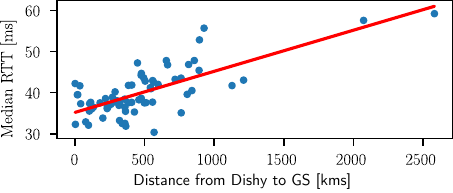}
        \vspace*{-1.5em}
	\caption{\label{fig:ripe_abs_dist} Correlation between Starlink bent-pipe latency and Dishy-GS distance. Red line denotes linear regression fit.}
    \vspace*{-1.5em}
\end{figure}

\headline{Impact of Ground Infrastructure}
% Our global analysis hinted at a correlation between ground station and PoP availability on Starlink network performance. 
%
% We extend our investigation by correlating the distance between Starlink user to ground station (GS) on bent-pipe latencies.
We extend our analysis by exploring the correlation between the distance from Starlink users to the GS and bent-pipe latencies.
Recall that we rely on crowdsourced data~\cite{starlink-gs-pop-unofficial} for geolocating Starlink ground infrastructure since these are not officially publicly disclosed.
% necessitating reliance .
%
We deduce through our \texttt{traceroute}s that Starlink directs its subscribers to the nearest GS relative to the PoP, as the GS $\leftrightarrow$ PoP latencies are $\approx$~5~ms (almost) globally (see \Cref{fig:ripe_tracert_gs_pop_global_abs} in \Cref{app:global} -- sole exceptions being US and Canada with 7--8~ms, likely due to abundant availability of GSs and PoPs causing routing complexities).
%
% We corroborate the GS locations for RIPE Atlas Starlink probes by reverse-geolocating based on GS-PoP latency and aligning the findings with the crowdsourced data (see \Cref{app:id-pop-location} for our PoP geolocation methodology).
%
\Cref{fig:ripe_abs_dist} correlates the reported bent-pipe latency with the terminal $\leftrightarrow$ GS distance.
%
% plotted against the distances of all probes to the ground stations they connect to. 
%
% We filter out all instances where the GS and PoP are co-located (i.e. RTT difference $\leq$ 1ms between them). \Cref{app:id-pop-location} discusses how we have identified the respective POP locations from our traceroute measurements. 
%
Each point in the plot denotes at least 1000 measurements.
%
% The red line shows the regression fit. % mentioned in figure caption
%
% we plot the CDF of the last-mile latency according to this distance group. 
We observe that
% a directly proportional relationship as 
bent-pipe latencies tend to increase with increasing distance to the GS. 
%
% While most probes attach to GSs within 1000~km 
Furthermore, we find that the predominant distance between GS and the user terminal is $\leq$ 1200~km, which is also the approximate coverage area width of a single satellite from 500~km altitude~\cite{cakaj2021parameters} -- suggesting that these connections are likely using direct bent-pipe, either without or with short ISL paths.
%
% The outlier at 500~km is a probe in the north of France connecting to a GS and PoP in Germany.
% %
% We are unsure of the reason for such inflated bent-pipe latency, but we suspect non-optimal ISL routing as potential culprit~\cite{izhikevich2023democratizing,tanveer2023making}.
%
%
Few terminals, specifically in Reunion, Falkland and the Virgin Islands, connect to GSs significantly farther away, possible only via long ISL chains, 
% the impact of 
which we analyze further as a case study.
%
% The bent-pipe latencies for these 
% (understandable due to a longer ``bent-pipe'').
%
% We also analyze the C\textsubscript{v} across different distance groups, which also appear to be similar (not shown).
%
% The result is understandable since the speed-of-light 
% As such, we conclude that from our measurements, the distance between and the ground station does not appear to impact the last-mile latency.
%  suggesting that the variability in last-mile latency does not vary with distance to the ground station.
% last-mile latency where a probe is within a distance of 101 km is slightly lower as compared to greater distances to the ground station. We consider only probes which are being served by satellites in the same inclination angle of 53$\degree$, as we expect the latency characteristics might differ for probes in different latitudes which are served by satellites in different inclination angles [ref]. We notice that the C\textsubscript{v} illustrated in figures [ref] and [ref] are also quite similar for all the groups, 
% thus inferring the variability in last-mile latency does not vary with distance to the ground station.

\begin{figure}[t!]
    \includegraphics[width=\columnwidth]{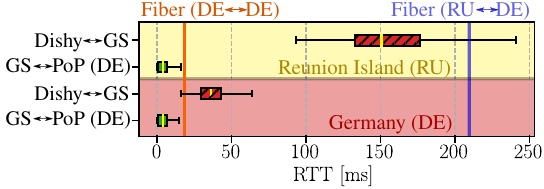}
    \vspace*{-2.5em}
\caption{\label{fig:isl_v_non_isl} Bent-pipe RTT segments from Reunion Island (yellow) vs. Germany (red) connecting to Germany PoP. Vertical lines show latency over Atlas probes connected via fiber from both locations to the Frankfurt server (PoP location).}
    \vspace*{-2em}
\end{figure}

\headline{Case Study: Reunion Island}
The majority of Starlink satellites (starting from v1.5 deployed in 2021) are equipped with ISLs~\cite{starlink-isl}, and reports from SpaceX suggest active utilization of these links~\cite{starlink-isl-use}.
Recent studies also agree with the use of ISLs~\cite{izhikevich2023democratizing}, but point out inefficiencies in space routing~\cite{tanveer2023making, leoselfdriving-mobicom23}.
Nonetheless, the invisibility of satellite hops in \texttt{traceroute}s poses a challenge in accurately assessing the use or impact of ISLs.
%
% Starlink currently supports coverage for maritime services from July 2022 onwards \footnote{https://www.businessinsider.com/spacex-starlink-internet-boats-sea-rvs-vehicles-elon-musk-2022-7}. This has been made possible with their newer launched satellites capable of communicating with each other over inter-satellite links (ISLs). \Cref{fig:starlink-bentpipe} illustrates how the last-mile path differs (user to GS) in the case of a one-hop relay versus with usage of ISLs. This difference can lead to performance variations for the last-mile in terms of latency. The ISL hops are unfortunately not visible in the traceroute measurements from the RIPE Atlas probes, and thus indistinguishable. However, further analysis on which PoP locations a probe is connecting to (see \Cref{app:id-pop-location}), 
As such, we focus on a probe in Reunion Island (RU), which connects to the Internet via Frankfurt PoP ($\approx$ 9000~km).
%
% uncovered the usage of the Frankfurt PoP location for a  probe situated in Reunion Island (RU) (Africa). 
\Cref{fig:isl_v_non_isl} segments the bent-pipe RTT between the user terminal (Dishy) to GS (non-terrestrial), and from GS to the PoP (terrestrial).
For comparison, we also plot the RTTs from a probe within Germany (DE) connecting to the same PoP ($\approx$ 500~km, in red).
The vertical lines represent the median RTT over terrestrial infrastructure from both probe locations to the PoP.
Firstly, we observe minimal GS $\leftrightarrow$ PoP latency for both locations, verifying that the RU satellite link is using ISLs.
Secondly, in RU, Starlink shows significant latency improvement over fiber ($\approx$ 60~ms). This is because the island has limited connectivity with two submarine cables routing traffic 10,000~km away, either in Asia or South America~\cite{noordally2016aintec}.
Starlink provides a better option by avoiding the terrestrial route altogether, directly connecting RU users to the dense backbone infrastructure in EU~\cite{ching2015arewe}.
However, since the bent-pipe incurs at least 30--40~ms latency in the best-case, Starlink is less attractive in regions with robust terrestrial network infrastructure (also evident from the DE probe where fiber achieves better latencies). 
% The probe in DE, on the other hand, experiences a higher latency ($\approx$ 15~ms) than the terrestrial connection, as the satellite link is likely to be routed via ISLs to a GS in the US, and then back to the PoP in Frankfurt.
% for the RU probe to a probe in Germany. Both of them connect to the Frankfurt PoP.  The median RTT value for GS<->PoP path is close to 4ms for RU. As the RU probe connects to the Frankfurt PoP, hence this must be connecting to a ground station within 400kms (1ms of RTT $\approx$ 100Kms of fibre cable). This would not be possible if a connection was made over a one-hop satellite relay, but rather with help of ISLs.
% The ISL routing can itself also be indirect and dynamic depending on the satellite constellations at any given point of time. Considering a distance of 9.1 Kms between RU and Frankfurt, the RTT values for "UE<->GS" would be around 60 ms. However, the measurements show them to be varying from 100 to 250 ms, which hints that high variability might be present in ISL communication. This might be possible due to indirect routing or availability of ISL usage at a given time.

\begin{figure}[t!]
	\includegraphics[width=0.9\columnwidth]{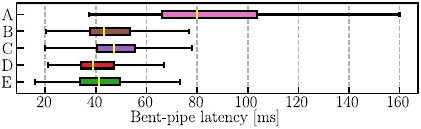}
    \vspace*{-1.5em}
	\caption{\label{fig:ripe_abs_cdf_lattitude}
    Bent-pipe latencies for ``A'' (in Alaska) covered by the 70$\degree$ and 97.6$\degree$ while the rest (Sweden ``B'', Canada ``C'', UK ``D'', and Germany ``E'') are also covered by 53$\degree$.}
    \vspace*{-2em}
\end{figure}

% \smallskip
% \noindent\textbf
\headline{Impact of Serving Orbit}
Recall that the majority of Starlink satellites are deployed in the 53$\degree$ inclination (see  \Cref{tbl:starlink_orbits} in \Cref{app:orbits}).
Consequently, network performance for clients located outside this orbit's range may vary widely as they are serviced by fewer satellites in 70$\degree$ and 97.6$\degree$ orbits.
%
% Finally, we compare the last-mile latencies for probes located within 53$\degree$ N and those which lie above this altitude. These two groups are serviced by satellites deployed in different inclination angles, and served by different number of satellites [ref.]. 
% To investigate this, in 
\Cref{fig:ripe_abs_cdf_lattitude} contrasts the bent-pipe latencies of probe in Alaska (61.5685N, 149.0125W) [``A''] to probes within 53$\degree$ orbit.
% (deployed in Poland [``B''], Canada [``C''], UK [``D''], and Germany [``E''])
%
% As discussed in \Cref{sec:background}, Dishys positioned outside of the 53$\degree$ inclination orbits will connect to the limited number of satellites in the 70$\degree$ and 97.6$\degree$ orbits.
%
% Note that there is a possibility that the terminal in Alaska is \emph{also} able to connect to satellites in 53$\degree$ inclination orbit since recent high-performance dishy's from Starlink claim to see ``35\% more sky'' that traditional releases~\textcolor{red}{[cite]}.
% We cannot refute this possibility since we do not have access to the dish or know of its detailed specifications.
% RIPE Atlas user data entry for the probe lacks any detad specification of the dishy, which is essenial to ensure this claim.
%
Despite dense GS availability, the bent-pipe latencies for Alaska are significantly higher ($\approx$ 2$\times$).
% compared to regions within 53$\degree$ orbits. 
% Our results show a clear distinction between Alaska and the other locations
% that confidently hints at different orbital connectivity from the probe.
% that confirms our assertion about the dish's connectivity. 
The Swedish probe [``B''] at 59.6395N is at the boundary of 53$\degree$ orbit but still exhibits comparable latency to Canada, UK, and Germany. 
% northern-most probe in our measurements having connectivity to the satellites in the 53$\degree$ orbits, also observes lower latency than the Alaska probe at 61.5685N connecting to only the 70$\degree$ and 97.6$\degree$ orbits.
% 
%
% As illustrated in figure [ref] the last-mile latency values are slightly higher for probe locations north of 53$\degree$. This could be due to the fact that the number of satellites in the inclination angles serving the higher latitudes are lower in number compared to the 53$\degree$ inclination angle.
% Unlike locations covered by 53$\degree$ also 
Furthermore, the Alaskan probe experiences intermittent connectivity, attributed to the infrequent passing of satellite clusters within the 70$\degree$ and 97.6$\degree$ orbits.
These findings indicate substantial discrepancies in Starlink's performance across geographical regions, which may evolve for the better as more satellites are launched in these orbits. 
% As we argue in \Cref{subsec:controlled}, interestingly the observed phenomenon in \Cref{fig:ripe_abs_cdf_lattitude} is not because that the satellite deployment has still not reached its intended target to cover all regions uniformly but is still linked to the variations in the density of Starlink's ground infrastructure (GSes and PoPs) in different regions as observed earlier in \Cref{sec:global}.
%
Nevertheless, we leverage the sparse availability of satellites at higher latitudes 
% regions currently provides us with an unique opportunity 
to further dissect the bent-pipe operations in \Cref{subsec:controlled}.

% \vspace*{-0.2em}
\begin{tcolorbox}[title=\textit{Takeaway \#3}, enhanced, breakable]
    The Starlink ``bent-pipe'' accounts for $\approx$ 40~ms latency globally.
    In certain cases with ISLs use, the latencies might escalate, yet still outshine traditional terrestrial networks when bridging remote regions.
    The satellite link yields stable latencies, provided that the client is served by the dense 53$\degree$ orbit.
    % Outside this region, latencies show a multi-fold increase, with intermittent connectivity due to the limited number of satellites in other inclinations.
  \end{tcolorbox}

\vspace*{-1em}
\subsection{Controlled Experiments} \label{subsec:controlled}

% We now investigate the cause
% reconfiguration intervals that caused 
% of periodic disruptions to real-time applications (\Cref{sec:cloud-gaming}). Specifically, we perform high-resolution measurements to gain insights into Starlink network operation.
% and scheduling of the Starlink network.

\begin{figure*}[!t]
    \centering
    \begin{subfigure}[htbp]{0.34\textwidth}
        \includegraphics[width=\textwidth]{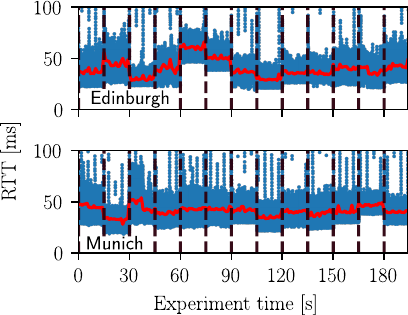}
        \phantomcaption
        \label{fig:last_mile_two_dish}
    \end{subfigure}
    \begin{subfigure}[htbp]{0.32\textwidth}
        \includegraphics[width=\textwidth]{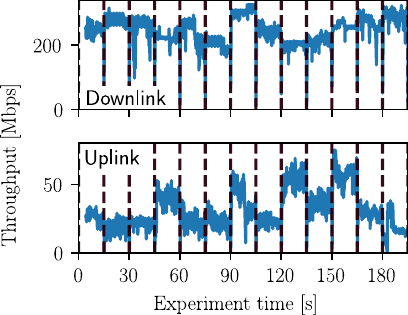}
        \phantomcaption
        \label{fig:last_mile_single_dish_throughput}
    \end{subfigure}
    \begin{subfigure}[htbp]{0.32\textwidth}
        \includegraphics[width=\textwidth]{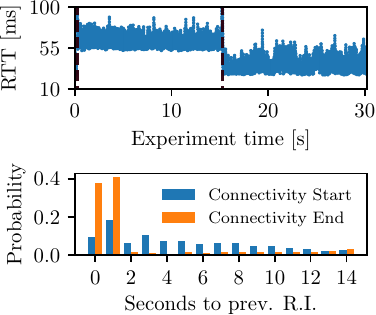}
        \phantomcaption
        \label{fig:last_mile_fov}
    \end{subfigure}
    \vspace*{-2.5em}
    \caption{(left, a) \texttt{iRTT} latencies with Dishys in two countries connected to different ground infrastructure; (middle, b) Maximum uplink and downlink throughput over a 195-second (13 interval) period; (right, c) (upper) RTTs for a connectivity window where the Dishy was connected to only a single satellite; (lower) Probability distribution of the time between the connectivity window start / end and the previous reconfiguration interval (RI). Vertical dashed lines show Starlink reconfiguration intervals.}
    \vspace*{-1.2em}
\end{figure*}

\headline{Global Scheduling}
We performed simultaneous \texttt{iRTT} measurements from terminals in Edinburgh (UK) and Munich (DE).
Note that the countries are sufficiently geographically removed that both cannot be connected to the same serving satellite and are assigned different PoPs. 
%
% We also verify that both terminals are assigned different PoPs located within their country.
%
The resulting RTTs, shown in Figure~\ref{fig:last_mile_two_dish}, vary consistently, being comparatively stable within each Starlink reconfiguration interval but potentially changing between intervals.
% Additionally, the reconfiguration intervals for both vantage points align, ruling out per-Dishy or per-satellite intervals, which demonstrates that Starlink is globally coordinated.
Moreover, the time-wise alignment of reconfiguration intervals for both vantage points indicates that Starlink operates on a globally coordinated schedule, rather than on a per-Dishy or per-satellite basis.
These results are in line with other recent studies~\cite{tanveer2023making}, which also hint that Starlink utilizes a global network controller.
% which demonstrates that Starlink is a truly globally-coordinated network.
%
% \headline{Throughput Impacts}
%
Previous studies~\cite{garcia2023multi} have noticed drops in downlink throughput every 15~s
but have not correlated these with the reconfiguration intervals.
We also observe throughput drops on both downlink and uplink, shown in Figure~\ref{fig:last_mile_single_dish_throughput}, that occur at the reconfiguration interval boundaries.
Similar to the RTT, the throughput typically remains relatively consistent within an interval, but fluctuates between intervals.
% but can experience sudden changes between interval transitions.
%
These results corroborate the periodic degradations observed in \Cref{sec:cloud-gaming}.
% our real-time application experiments.
% Beyond this, we identify similar characteristics between RTT and throughput:
% that the throughput typically remains relatively consistent within an interval, but can experience sudden changes between intervals.

\headline{Disproving Satellite Handoff Hypothesis}
Previous works have suggested satellite or beam changes at
% that have identified changes in RTT and throughput at 
reconfiguration interval boundaries to be the root-cause of network degradation~\cite{tanveer2023making, garcia2023multi, fcc-15s}. 
% Indeed, the FCC filing~\cite{fcc-15s} supports this hypothesis, describing the intervals as ``handing off connections between satellites''. 
To investigate this hypothesis, we deliberately obstructed the field-of-view of our UK terminal to prevent it from connecting to the dense 53\degree\ shell (see \cref{sec:methodology:controlled} for details). 
The restriction curtailed the number of candidate (potentially connectable) satellites to 13\%, resulting in intermittent connectivity.
%
% This limitation led to intermittent connectivity, characterized by brief connectivity windows with long service downtimes. 
%
By synchronizing the timings of each connectivity window with the overhead positions of candidate satellites (from CelesTrak~\cite{celestrak2023} and other sources~\cite{starlinksx2023}), we identify several windows where the terminal can be served by only a single satellite.
Figure \ref{fig:last_mile_fov} (upper) shows RTTs from such a window.
The significant RTT variance between intervals invalidates the hypothesis that the RTT changes are caused by satellite handovers (no handoffs are possible with single satellite in field-of-view).
%  a single candidate satellite during the observed period, leaving no room for hand-off occurrences).
%The variance in RTTs between intervals arises due to another factor in the network.
%
% Separately, we perform the same experiment but focus on (both uplink and downlink) throughput.
%
Similar to RTT, we also witness uplink and downlink throughput drops at interval boundaries even when single candidate satellite is visible.
% Therefore these drops are not explained solely by hand-offs between satellites.

\headline{Scheduling Updates}
%
% To gain deeper insight into the routing allocation system used by Starlink, 
Figure~\ref{fig:last_mile_fov} (lower) shows the distribution of start and end times of the connectivity windows during our restricted field-of-view experiments.
% Specifically, we focused on the distribution of the start and end of the connectivity windows across the reconfiguration intervals, shown in .
We observed a strong correlation between connectivity end times and reconfiguration interval (RI) boundary, which is not seen with start times\footnote{The fact that many appear to end 1s after the boundary is an artifact of the limited (per-second) granularity of the gRPC data and that the gRPC timestamps originate from the client making the gRPC requests rather than the user terminal.}.
The result hints at internal network scheduling changes at reconfiguration interval boundaries, i.e., Starlink assigns its terminals new satellites (or frequencies) every 15s.
We hypothesize that with an obstructed view, the scheduler cannot find better alternatives in the 70$\degree$ and 97.6$\degree$ orbits, resulting in connectivity loss at the end of the window.
% which results in connectivity loss  
% We observe that connectivity is approximately equally likely to begin at any point within a reconfiguration interval, while it is statistically very likely (>70\%) to end at the reconfiguration interval boundary.

\headline{Analysis Summary}
Putting together our various observations, we theorize that Starlink relies on a global scheduler that re-allocates the user $\leftrightarrow$ satellite(s) $\leftrightarrow$ GS path every 15s.
An FCC filing from Starlink implies this behavior~\cite{fcc-15s} and recent studies also suggest that the LEO operator performs periodic load balancing at reconfiguration boundaries, reconnecting all active clients to satellites~\cite{tanveer2023making, izhikevich2023democratizing}.
The theory also explains our observed RTT and throughput changes when only a single candidate satellite is in view. It is plausible that Starlink may have rescheduled the terminal to the same satellite but with reallocated frequency and routing resources.
Regardless, these reconfigurations result in brief sub-second connection disruptions, which may become more noticeable at the application-layer as the number of subscribers on the network increases over time.

% As the demands for real-time applications increase, we expect the impact of the reconfiguration interval to become significantly more apparent in the end-user performan

% \headline{Summary Analysis}
% %
% Considering our findings in aggregate, we speculate that Starlink globally assigns routes on each connectivity interval.
% Each route including a serving satellite, ground station, PoP, and route through the Starlink network.
% The drops in throughput would be caused by the need to momentarily switch between routes (even if reconnecting to the same satellite), and would affect all traffic -- both downlink and uplink.
% The variations in RTT and throughput between intervals would arise from different paths in the new route, while the more consistent performance within an interval is due to the same route.
% The fact that the effects of the interval boundary are observed even when continuously connected to the same satellite could be explained by the new route containing the same satellite, but a different path beyond that.
% Finally, the difference between the probability distributions (Figure~\ref{fig:last_mile_fov}) could be caused by the (roughly evenly distributed) time it takes the dish to receive a schedule from an available satellite, identify the satellite and connect, and the end of connectivity when it is not scheduled a route or scheduled to a satellite out of view due to the restricted field-of-view (likely to occur on interval boundary).

% \vspace*{-0.2em}
\begin{tcolorbox}[title=\textit{Takeaway \#4}, enhanced, breakable]
    Starlink uses 15s-long reconfiguration intervals to globally schedule and manage the network. Such intervals cause latency/throughput variations at the interval boundaries. Hand-offs between satellites are not the cause of these effects. Indeed, our findings hint at a scheduling system reallocating resources for connections once every reconfiguration interval.
\end{tcolorbox}

\vspace*{-1em}

\section{Related Work} \label{sec:relatedwork}

% \noindent
% \textbf{Theoretical and simulation based studies.}
%
LEO satellites have become a subject of extensive research in recent years, with a particular focus on advancing the performance of various systems and technologies.
Starlink, the posterchild of LEO networks, continues to grow in its maturity and reach with $>$ 2M subscribers as of September 2023~\cite{starlink-subs}.
%
% However, there is little public knowledge on the performance of Starlink at a global scale~\cite{satnetlab}, with only a few works shining light on its end-to-end operation.
% However, there have only been limited explorations measuring Starlink's performance so far.
Despite its growing popularity, there has been limited exploration into measuring Starlink's performance so far.
Existing studies either have a narrow scope, employing only a few vantage points~\cite{michel2022, garcia2023multi, lopez2022} or focus on broad application-level operation~\cite{kassem2022, zhao2023real} without investigating root-causes. 
% a few recent investigations have attempted to uncover its perfomance from different viewpoints.
%
% \citet{kassem2022} investigated web browsing performance with Starlink through their browser-based crowdsourced measurements.
% plugin that periodically load top-200 webesites in the Tranco list. 
% 
% The authors observed significant variability in page load times (PLT) and throughput over Starlink compared to terrestrial access alternatives.
% They attributed it to the ``bent-pipe'' last-mile access and external factors such as weather and location.
%
% The authors also conduct traditional \texttt{iperf} and \texttt{ping} measurements using volunteer Starlink terminals in three locations (east-coast USA, UK and Spain) towards nearest Google cloud server.
% %
% The authors uncover similarly high variation in throughput with periodic packet loss which they attribute to satellite handovers.
%
% \citet{michel2022}, on the other hand, investigated the transport layer performance over Starlink (with a particular focus on QUIC) and compared it to GEO SatCom networks.
%
% The authors also uncovered similar baseline latencies as \cite{kassem2022} ($\approx$ 20~ms) %with significant variability and bursty packet loss under load.
%
% The authors also verify 
% and showed that LEO Starlink significantly outperforms GEO SatCom -- achieving 70--80\% application speedups.
%
% \citet{ma2022network} embarked on a journey across Canada with four dishes to scrutinize various factors, such as temperature and weather, that might influence Starlink's performance.
%
Few studies have looked into the mobile behavior of Starlink and compared it to terrestrial cellular carriers~\cite{ma2022network,leoVcellular-conext23}.
%
% They test the dish's mobile behavior and perform simple video streaming tests during which they find correlations between close-by Dishys.
%

% \citet{lopez2022} reported superior latency and upload throughput over a 5G connection compared to Starlink, while the measured download speeds were higher with Starlink.
% Their measurements were, however, limited to a single test location.
% Several works have focused on shining light on the Starlink black-box network operation, e.g.
A few endeavors have attempted to unveil the operations of Starlink's black-box network. 
\citet{pan2023measuring} revealed the operator's internal network topology from \texttt{traceroute}s, whereas \citet{tanveer2023making} spotlighted a potential global network controller. 
% that allocates satellites to user terminals at 15-second intervals.  
% Compared to the prior Starlink-focused measurement studies discussed above, our measurement-based analysis is not only significantly larger in scope (with measurement vantage points spanning the globe) and scale (in terms of the number of measurement samples) but also has a deeper focus on revealing the ``bent-pipe'' behavior.
The absence of global measurement sites poses a predominant challenge hampering a comprehensive understanding of Starlink's performance. As we show in this work, Starlink's performance varies geographically due to differing internal configurations and ground infrastructure availability.
Some researchers have devised innovative methods to combat this.
For example, \citet{izhikevich2023democratizing} conducted measurements towards exposed services behind the Starlink user terminal, while \citet{taneja2023viewing} mined social media platforms like Reddit to gauge the LEO network's performance.
Our study not only corroborates and extends existing findings but also stands as the most extensive examination to date.
% We achieve this in several ways.
%
% In particular, we provide a \emph{global} view of the Starlink network performance from the end-user perspective via 
% Notably, we offer detailed insights from 34 countries and leverage
% 19.2~M crowdsourced M-Lab measurements, 2.9~M active RIPE Atlas measurements, and two controlled terminals connecting to different Starlink orbits.
Our approach -- anchored in detailed insights from 34 countries, leveraging 19.2 million crowdsourced M-Lab measurements, 1.8 million active RIPE Atlas measurements, and two controlled terminals connecting to different Starlink orbits -- provides a deeper understanding of the Starlink ``bent-pipe'' and overall performance.

\section{Conclusions} \label{sec:conclusion}

Despite its potential as a ``global ISP'' capable of challenging the state of global Internet connectivity, there have been limited performance evaluations of Starlink to date.
We conducted a multi-faceted investigation of Starlink, providing insights from a global perspective down to internal network operations.
Globally, our analysis showed that Starlink is comparable to cellular for supporting real-time applications (in our case Zoom and Luna cloud gaming), though this varies based on proximity to ground infrastructure.
Our case study shows Starlink inter-satellite connections helping remote users achieve better Internet service than terrestrial networks.
However, at sub-second granularity, Starlink exhibits performance variations, likely due to periodic internal network reconfigurations at 15s intervals.
We find that the reconfigurations are synchronized globally and are not caused by satellite handovers. 
As such, this first-of-its-kind study is a step towards a clearer understanding of Starlink’s operations and performance as it continues to evolve.
% Despite only being available to the public for less than three years, Starlink has demonstrated the potential for LEO networks to provide global, low-latency connectivity.
% This paper shows that a LEO satellite constellation alone is not a ``silver bullet''. As things stand, two key factors affecting the network performance are investment in ground infrastructure (nearness of GSes and PoPs to terminals) and network management.
% The former forms the largest contributor to observed RTTs due to the bent-pipe, while the latter introduces artifacts that increase the variability of the connection.
% While we observe that real-time applications are not currently affected by such artifacts, and indeed operate with similar performance to a terrestrial network, growing demands from applications that require both high-bandwidth and real-time could strain the network beyond its current limits.
% It remains an open question whether the planned advancements in the network, in particular the inter-satellite links and the deployment of more satellites in the polar inclinations, will be an effective solution as the requirements of applications and the number of users of Starlink both increase.

% \pagebreak

% \clearpage

\begin{acks}
 We thank the anonymous reviewers for their valuable feedback.
 This work was supported by the German Federal Ministry of Education and Research joint project 6G-life (16KISK002) and the German Federal Ministry for Digital and Transport project 5G-COMPASS (19OI22017A).
 We also thank the Starlink mailing list~\cite{starlinkmailinglist} which motivated several investigations within this work.
 Finally, we thank the Measurement Lab and RIPE Atlas platforms for providing us access to their dataset and infrastructure.

\end{acks}

%% Reference section
% \pagebreak
% \nocite{*}
\bibliographystyle{ACM-Reference-Format}
\bibliography{library}
% \printbibliography{}

\pagebreak

\clearpage

\appendix

\section{Starlink Orbital Information}\label{app:orbits}
\vspace*{-0.5em}

%% TODO: Cite properly
% source: https://planet4589.org/space/con/star/stats.html (Accessed: 2023-05-04)
\begin{table}[th]
\centering
\begin{adjustbox}{width=\columnwidth}
\begin{tabular}{@{}cccccc@{}}
\toprule
\multirow{2}{*}{\textbf{\begin{tabular}[c]{@{}c@{}}Inclination\\ angle\end{tabular}}} &
  \multirow{2}{*}{\textbf{\# Planes}} &
  \multirow{2}{*}{\textbf{\begin{tabular}[c]{@{}c@{}}Altitude\\ {[}km{]}\end{tabular}}} &
  \multicolumn{3}{c}{\textbf{\# Satellites}} \\
    \cmidrule(l){4-6}
      &    &     & \textbf{\begin{tabular}[c]{@{}c@{}}In\\ Position\end{tabular}} & \textbf{Launched} & \textbf{Filed~\cite{fcc-1, fcc-2}} \\
  \midrule
53°   & 72 & 550 & 1401                                                           & 1665              & 1584           \\
53.2° & 72 & 540 & 1542                                                           & 1637              & 1584           \\
70°   & 36 & 570 & 301                                                             & 408               & 720            \\
97.6° & 10 & 560 & 230                                                            & 230               & 508           \\
\bottomrule
\end{tabular}
\end{adjustbox}
    \caption{\label{tbl:starlink_orbits} Starlink orbital shell design and number of operational satellites as of October 2023~\cite{planet4589}.}
    \vspace*{-2em}
\end{table}

% This table does not include 2\textsuperscript{nd} generation Starlink satellites~\cite{starlink-v2-0}.

% \begin{figure}[htbp]
%     \centering
%     \includegraphics[width=0.35\textwidth]{figures/orbit_render.png}
%     \caption{\label{fig:mlab_shells_render} 3D rendering of Starlink's orbital shells. The 3D representation accompanies the flattened world map in \Cref{fig:mlab_shells}.}
% \end{figure}

Starlink and other emerging LEO satellite constellations, such as OneWeb and Kuiper~\cite{oneweb,kuiper}, are termed \textit{megaconstellations}
since they combine multiple orbital shells compared to single shell systems like Iridium~\cite{iridium}.
% Early satellite constellations for telephone services only consisted of a single shell~\cite{iridium}.
%
\Cref{tbl:starlink_orbits} details the number of satellites and their altitude in Starlink's orbital shells.
% The eccentricity parameters of Starlink orbits are negligible due to the satellites' positions in Low Earth Orbit (LEO).
While discussing Starlink's constellation design, we simplify the orbit into circular orbits.
%
% \Cref{fig:mlab_shells_render} shows how orbits of different orbital inclinations compare to each other. An \textit{orbital shell} refers to the collection of satellite orbits sharing the same inclination and altitude.

\section{Data Center Endpoints} \label{app:cloud-dcs}

\begin{table}[p]
	%\begin{minipage}[b]{0.58\linewidth}
	\centering
	\begin{adjustbox}{width=0.9\columnwidth,center}
		\begin{tabular}{@{}lrrrrrr@{}}
			\toprule
			\multirow{2}{*}{}                & \multicolumn{6}{c}{\textbf{Data centers per continent}}                         \\ \cmidrule(l){2-7}
											 & EU          & NA          & SA         & AS          & AF         & OC         \\ \midrule
			\textbf{Amazon EC2 (AMZN)}       & 6           & 6           & 1          & 6           & 1          & 1          \\
			\textbf{Google Cloud (GCP)}      & 6           & 10          & 1          & 8           & -          & 1          \\
			\textbf{Microsoft Azure (MSFT)}  & 14          & 9           & 1          & 10          & 2          & 3          \\
			\textbf{Digital Ocean (DO)}      & 4           & 6           & -          & 1           & -          & -          \\
			\textbf{Alibaba (BABA)}          & 2           & 1           & -          & 2           & -          & 1          \\
			\textbf{Amazon Lightsail (LTSL)} & 3           & 2           & -          & 2           & -          & 1          \\
			\textbf{Oracle (ORCL)}           & 4           & 4           & 1          & 7           & -          & 2          \\ \midrule
			\textbf{Total}                   & \textbf{39} & \textbf{38} & \textbf{4} & \textbf{36} & \textbf{3} & \textbf{9} \\ \bottomrule
			\end{tabular}%
	\end{adjustbox}
	\caption{\label{tab:cloud-dc-breakup}Global density of data center endpoints used for RIPE Atlas measurements (\cref{sec:last-mile}).}
	% \vspace*{-1em}
	\end{table}

 \Cref{tab:cloud-dc-breakup} shows the distribution of our cloud data center endpoints by cloud provider and deployed continent.
Each endpoint is a VM in a compute-capable cloud data center location.
Our selection is influenced by previous studies that have found that significant end-to-end performance differences may appear while measuring different cloud networks due to private WANs, peering agreements, etc.~\cite{dang2021, corneoRIPE}.
We believe that our comprehensive endpoint selection reduces potential biases due to Internet traffic steering.
%  that may affect our aggregated analysis.

\section{Global Starlink Performance}\label{app:global}

\begin{figure}[p]
	\centering
        \includegraphics[width=\columnwidth]{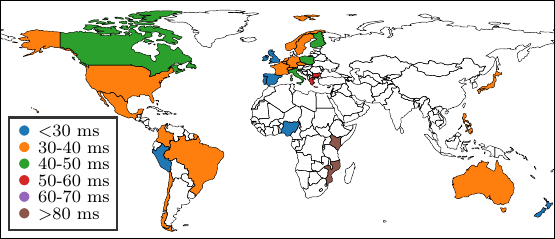}
        \caption{\label{fig:mlab_heatmap_starlink_latency}  Median Starlink last-mile latencies from M-lab. We subtract the M-lab server $\leftrightarrow$ PoP latency in reverse tracroute from overall measured min RTT.}
\end{figure}

\begin{figure}[p]
	\centering
        \includegraphics[width=\columnwidth]{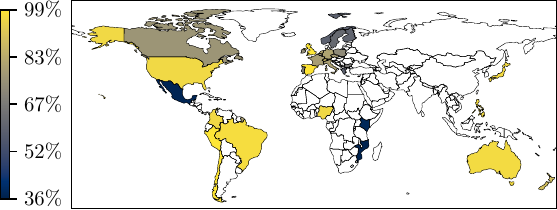}
        \caption{\label{fig:mlab_heatmap_starlink_latency_fraction}  Fraction of the latency, that is estimated to be over the satellite link by dividing the latency of \Cref{fig:mlab_heatmap_starlink_latency} with its overall latency.}
\end{figure}

\begin{figure}[p]
	\centering
        \includegraphics[width=\columnwidth]{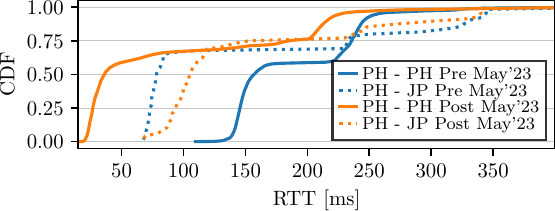}
        \caption{\label{fig:mlab_cdf_minrtt_phjp} Comparing median minRTT from tests originating in Manila that targeted M-Lab servers in the Philippines and in Japan. The results are discussed in \Cref{sec:global}.}
\end{figure}

\begin{figure*}[!t]
  \begin{subfigure}[htbp]{\textwidth}
      \includegraphics[width=\textwidth]{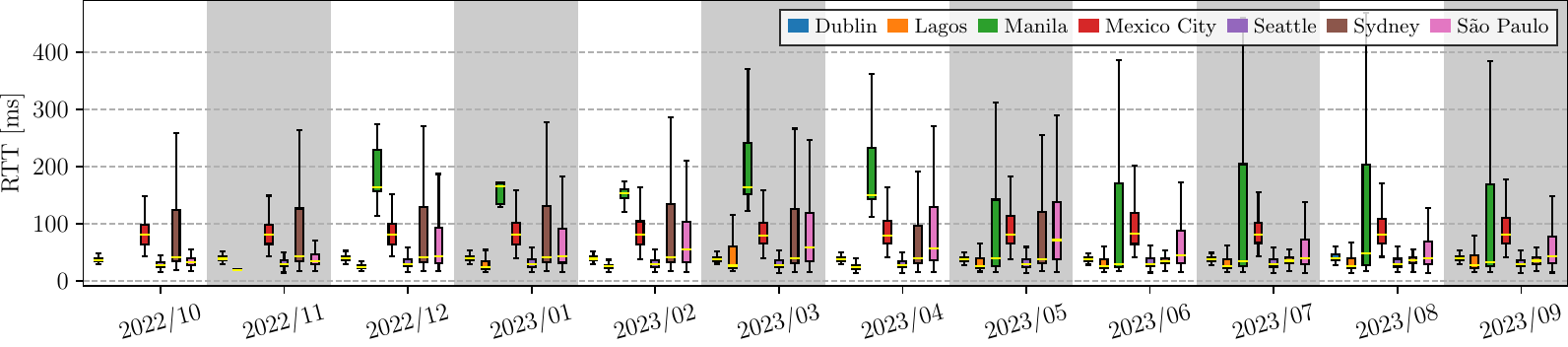}
      \vspace*{-2em}
      \caption{\label{fig:mlab_temporal_2022-2023_minrtt} minimum RTT}
  \end{subfigure}
  % \vspace*{-1em}
  \begin{subfigure}[htbp]{\textwidth}
      \includegraphics[width=\textwidth]{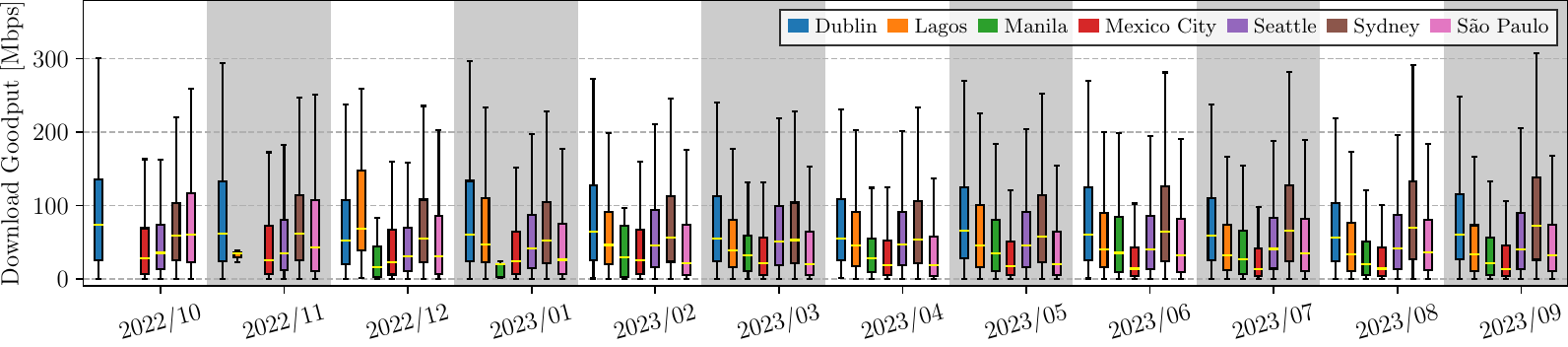}
      \vspace*{-2em}
      \caption{\label{fig:mlab_temporal_2022-2023_dl_goodput} Download Goodput}
  \end{subfigure}
  % \vspace*{-0.5em}
  \begin{subfigure}[htbp]{\textwidth}
      \includegraphics[width=\textwidth]{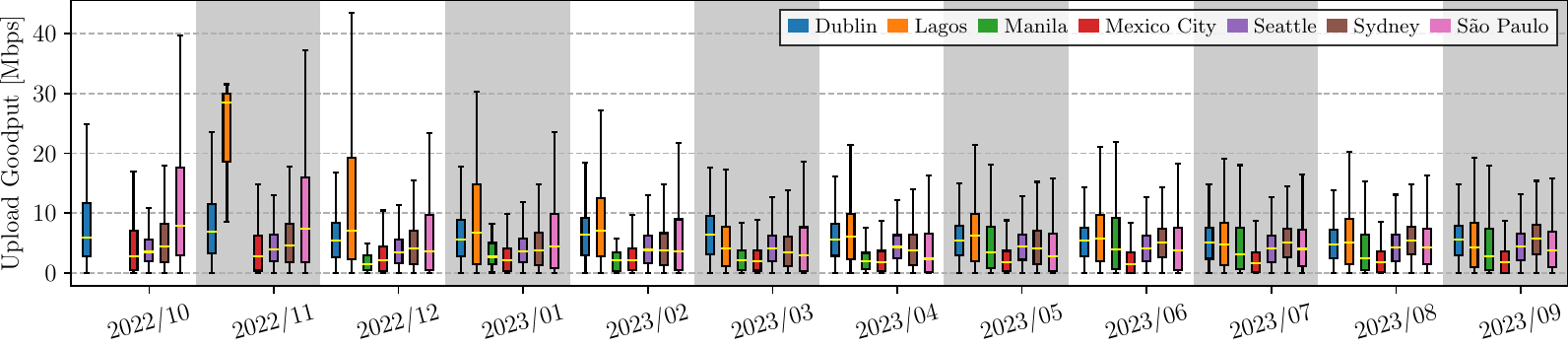}
      \vspace*{-2em}
      \caption{\label{fig:mlab_temporal_2022-2023_ul_goodput} Upload Goodput}
  \end{subfigure}
  \vspace*{-1em}
  \caption{\label{fig:mlab_temporal_global}Evolution of Starlink aggregate goodput ((a), (b)) and minimum RTT (c) during download measurements from cities in South America, North America, Europe, and Australia in the last 12 months.}
\end{figure*}

\begin{figure*}[!htb]
	\begin{minipage}{0.325\linewidth}
            \includegraphics[width=0.9\columnwidth]{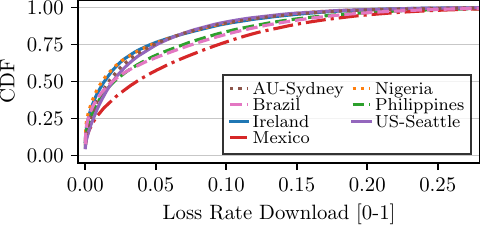}
            \caption{\label{fig:mlab_cdf_lossrate-down_global}
              TCP losses during M-Lab downloads from selected cities globally.
              % The results are set into context in \Cref{sec:global}.
              }
	\end{minipage}%
	\hfill
	\begin{minipage}{0.65\linewidth}
            \begin{subfigure}[htbp]{0.5\textwidth}
                \includegraphics[width=\textwidth]{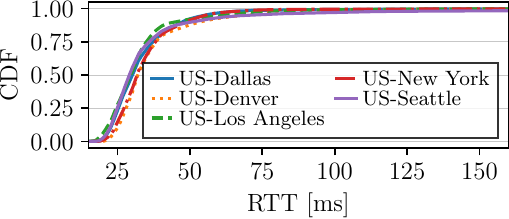}
                \vspace*{-1.8em}
                \caption{\label{fig:mlab_cdf_minrtt_na} North America}
            \end{subfigure}
            % \begin{subfigure}[htbp]{0.24\textwidth}
            %   \includegraphics[width=\textwidth]{figures/mlab/state_2023-10-01/mlab_cdf_minrtt_eucities_stripped.pdf}
            %   \caption{\label{fig:mlab_cdf_minrtt_eu} Europe}
            % \end{subfigure}
            % \begin{subfigure}[htbp]{0.24\textwidth}
            %   \includegraphics[width=\textwidth]{figures/mlab/state_2023-10-01/mlab_cdf_minrtt_5sacities_stripped.pdf}
            %   \caption{\label{fig:mlab_cdf_minrtt_sa} South America}
            % \end{subfigure}
            \begin{subfigure}[htbp]{0.5\textwidth}
                \includegraphics[width=\textwidth]{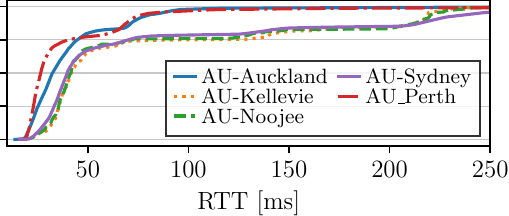}
                \vspace*{-1.8em}
                \caption{\label{fig:mlab_cdf_minrtt_au} Oceania}
            \end{subfigure}
            \vspace*{-1.5em}
            \caption{\label{fig:app:mlab_cdf_regions}
              The distribution of the minimum RTT (minRTT) during M-Lab measurements from selected cities in North America (a) and Oceania (b).
            %  Distributions of minimum reported RTTs during M-Lab measurements from selected cities in North America, Europe, South America and Oceania, respectively.
             }
	\end{minipage}
	\vspace{-1em}
\end{figure*}

% This section of the appendix provides supporting material for the global Starlink performance analysis presented in \Cref{sec:global}.
%
%\subsection{Evolution of Starlink Performance}
% As can be observed, the normal trend indicates a degradation in performance in terms of Goodput rate.
% This can be attributed to the higher traffic as more users join the network.
\Cref{fig:mlab_temporal_global} provides an overview of Starlink's performance over one year from selected cities in each continent.
% The plot gives an insight into the evolution of Starlink over time from a global perspective.
%
We observe decreasing goodputs over time, which can be largely attributed to an increase of Starlink users.
The RTT values, while relatively stable, are higher for countries outside of 
% which are not part of the 
major Starlunk operational areas. % of Starlink.
Sydney (AU) is a notable exception where RTT decreases over time.
\Cref{fig:mlab_cdf_minrtt_phjp} shows the median minRTT from Philippines to in-country vs Japanese endpoints.
The implications are discussed in \Cref{sec:global}.
\Cref{fig:mlab_cdf_lossrate-down_global} shows loss rates during M-Lab download measurements (see \Cref{sec:global}). %\Cref{sec:global:goodput}.
\Cref{fig:app:mlab_cdf_regions} complements \Cref{fig:mlab_cdf_regions} and shows the distribution of the minimum RTT (minRTT) during M-Lab measurements from selected cities in North America and Oceania.
% It complements  that depicts selected cities in Europe and South America.
Starlink's performance in North America varies slightly across different cities and all achieve lower latencies compared to rest of the globe.
In Oceania, tests from Auckland and Perth exhibit a similarly low minRTT due to PoPs and GSs near both cities.
Latencies in Sydney's has improved recently (2023/06) as shown in \Cref{fig:mlab_temporal_2022-2023_minrtt}.

\section{Global View of Bent-Pipe Operation}\label{app:last-mile:global}
\begin{figure}[h]
        \includegraphics[width=\columnwidth]{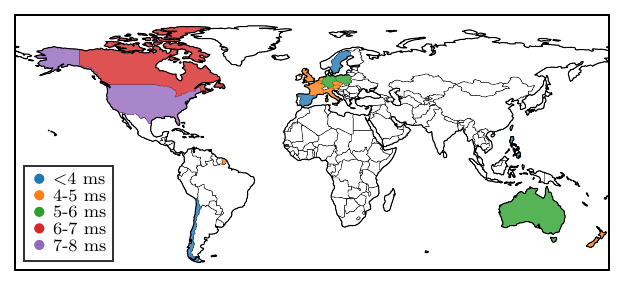}
        \vspace*{-2em}
        \caption{\label{fig:ripe_tracert_gs_pop_global_abs} Global GS $\leftrightarrow$ PoP latencies from Starlink RIPE Atlas probes.}
        \vspace*{-1.5em}
\end{figure}

% This section provides additional supporting material for the Starlink last-mile performance analysis presented in \Cref{subsec:ripeatlas}.
%
\Cref{fig:ripe_tracert_gs_pop_global_abs} augments 
% provides additional information about the 
our discussion on Starlink last-mile performance in \Cref{subsec:ripeatlas} and shows
% The figure shows the
GS $\leftrightarrow$ PoP latencies globally.
% latencies between Starlink ground stations (GSs) and Points of Presences (PoPs) on a world map grouped by country.
It is apparent that the latencies are similar all over the world ($\leq$~6~ms) except in North America ($\geq$~6~ms), likely due to dense deployment of GSs and PoPs in these regions (see \Cref{fig:mlab_shells}).

\section{Targeted Measurement Challenges}\label{app:last-mile:controlled}

% Some of the measurements required for Section 
We encountered several challenges during our controlled measurements over shielded terminal in \Cref{subsec:controlled} which we elucidate for researchers who plan to replicate our setup. 
% resulted inrequired collection of data during time when Dishy received only patchy connectivity. 
Specifically, both  \texttt{irtt} and  \texttt{iperf} could not handle interruptions well as they rely on single TCP connection which eventually times out. %and behaved unreliably.
% : \texttt{iperf} in particular relies on a separate TCP connection to act as the control plane. Both behaved unpredictably and unreliably on a link that has connectivity only for brief windows. Accordingly, for these experiments, 
To overcome this, we replaced \texttt{irtt} with periodic ICMP \texttt{ping} packets sent every 200~ms. 
% which we set to send only a single ICMP packet with a 200~ms timeout. We then ran it in a loop.
%
On the other hand, we manually controlled \texttt{iperf} to overcome connection drops. 
% We were unable to find a suitable replacement for  \texttt{iperf}, and therefore relied upon a manual approach. 
Specifically, we started \texttt{iperf} every time we detected the start of connectivity window (from \texttt{ping}s) and stopped the experiment upon interruption.
Before starting \texttt{iperf} at next window detection, we also restarted the server. 
% When we detected the start of a connectivity window  \texttt{iperf} was started. Once the connectivity window had passed, we stopped the experiment and restarted the  \texttt{iperf} server. 
% Restarting the  \texttt{iperf} server was necessary to ensure that subsequent  \texttt{iperf} tests could connect (\texttt{iperf3} permits only a single active connection to a server, and a loss of connectivity mid-way through an experiment can leave the server in a state where it believes an experiment is ongoing when it has in fact concluded). 
Automatically restarting the  \texttt{iperf} server at the end of each connectivity window was not possible because the Starlink-connected computer, now without an Internet connection, could not signal to the remote  \texttt{iperf} server.

An additional challenge caused by the interrupted nature of the connection became apparent in the analysis phase. The unstable connection prevented the clock on the machine connected to the Dishy from synchronising over NTP, resulting in it drifting by several seconds duration of the experiment setup. Accordingly, when the absolute timestamps of the recorded data were analysed, they were adjusted to account for the time slip. The gRPC data was collected from another machine that did not suffer from clock drift.

\end{document}